\documentclass[iop, apj, iopart]{emulateapj}
\usepackage{color}
\usepackage{adjustbox}
\usepackage[encapsulated]{CJK}
\usepackage{ucs}
\usepackage[utf8x]{inputenc}
\usepackage{longtable}
\usepackage[flushleft]{threeparttable}
\usepackage{array}
\usepackage{natbib}
\usepackage{multirow}
\usepackage{amsmath}
\newcolumntype{x}[1]{>{\centering\arraybackslash\hspace{0pt}}p{#1}}

\newenvironment{Contfigure*}{
\renewcommand{\thefigure}{\arabic{figure} (Cont.)}
\addtocounter{figure}{-1}
\begin{figure*}}{
\renewcommand{\thefigure}{\arabic{figure}}
\end{figure*}}

\newcommand{ \kms}{$\mathrm{km\,s^{-1}}$}

\begin{document}
\begin{CJK}{UTF8}{gbsn}

\title{The MOSDEF-LRIS Survey: Probing ISM/CGM Structure of Star-Forming Galaxies at $\lowercase{z}\sim2$ Using Rest-UV Spectroscopy}

\author{Xinnan Du (杜辛楠)\altaffilmark{1}, Alice E. Shapley\altaffilmark{2}, Michael W. Topping\altaffilmark{2,3}, Naveen A. Reddy\altaffilmark{1}, Ryan L. Sanders\altaffilmark{4,$\dagger$}, Alison L. Coil\altaffilmark{5}, Mariska Kriek\altaffilmark{6}, Bahram Mobasher\altaffilmark{1}, Brian Siana\altaffilmark{1}}

\altaffiltext{1}{Department of Physics and Astronomy, University of California, Riverside, 900 University Avenue, Riverside, CA 92521, USA}
\altaffiltext{2}{Department of Physics and Astronomy, University of California, Los Angeles, 430 Portola Plaza, Los Angeles, CA 90095, USA}
\altaffiltext{3}{Department of Astronomy / Steward Observatory, University of Arizona, 933 N Cherry Ave, Tucson, AZ 85721}
\altaffiltext{4}{Department of Physics and Astronomy, University of California, Davis, One Shields Avenue, Davis, CA 95616, USA}
\altaffiltext{5}{Center for Astrophysics and Space Sciences, Department of Physics, University of California, San Diego, 9500 Gilman Drive., La Jolla, CA 92093, USA}
\altaffiltext{6}{Astronomy Department, University of California at Berkeley, Berkeley, CA 94720, USA}
\altaffiltext{$\dagger$}{Hubble Fellow}

\slugcomment{Draft Version \today}

\shorttitle{ISM/CGM Structure}
\shortauthors{Du et. al}

\begin{abstract}

The complex structure of gas, metals, and dust in the interstellar and circumgalactic medium (ISM and CGM, respectively) in star-forming galaxies can be probed by Ly$\alpha$ emission and absorption, low-ionization interstellar (LIS) metal absorption, and dust reddening $E(B-V)$. We present a statistical analysis of the mutual correlations among Ly$\alpha$ equivalent width (EW$_{Ly\alpha}$), LIS equivalent width (EW$_{LIS}$), and $E(B-V)$ in a sample of 157 star-forming galaxies at $z\sim2.3$. With measurements 
obtained from individual, deep rest-UV spectra and spectral-energy distribution (SED) modeling, we find that the tightest correlation exists between EW$_{LIS}$ and $E(B-V)$, although correlations among all three parameters are statistically significant. These results signal a direct connection between dust and metal-enriched \textrm{H}\textsc{i} gas, and that they are
likely co-spatial. By comparing our results with the predictions of different ISM/CGM models, we favor a dusty ISM/CGM model where dust resides in \textrm{H}\textsc{i} gas clumps and Ly$\alpha$ photons escape through the low \textrm{H}\textsc{i} covering fraction/column density intra-clump medium. Finally, we investigate the factors that potentially contribute to the intrinsic scatter in the correlations studied in this work, including metallicity, outflow kinematics, Ly$\alpha$ production efficiency, and slit loss. Specifically, we find evidence that scatter in the relationship between EW$_{Ly\alpha}$ and  $E(B-V)$ reflects the variation in metal-to-\textrm{H}\textsc{i} covering fraction ratio as a function of metallicity, and the effects of outflows on the porosity of the ISM/CGM.
Future simulations incorporating star-formation feedback and the radiative transfer of Ly$\alpha$ photons will provide key constraints on the spatial distributions of neutral hydrogen gas and dust in the ISM/CGM structure.

\end{abstract}

\keywords{galaxies: high-redshift -- ultraviolet: galaxies -- ISM: structure}

\section{Introduction}
\label{sec:Intro}

Gas, metals, and dust in the interstellar and circumgalactic medium (ISM and CGM, respectively) play a key role in the evolution of galaxies.
Specifically, these components reflect the cycle of baryons through galaxies, and the processes of gas accretion, star formation, chemical enrichment, and stellar feedback -- which in turn alter the geometry and covering fraction of the ISM/CGM and determine the escape fraction of Lyman continuum (LyC) and Ly$\alpha$ radiation. However, a coherent picture regarding the ISM/CGM properties of star-forming galaxies has not yet been achieved, especially at high redshift. Many outstanding questions remain regarding the kinematics and spatial distributions of gas, metals, and dust. Therefore, detailed studies on the fundamental processes involving these baryonic components will enable us to gain a more comprehensive understanding of the ISM/CGM structure in star-forming galaxies and how it consequently affects the escape of radiation into the intergalactic medium (IGM) during the reionization epoch.

The rest-frame ultraviolet (UV) spectra of star-forming
galaxies provide a unique perspective on the physical properties of the ISM and CGM, the latter of which is typically defined as extending to the galaxy virial radius \citep{Tumlinson2017}. Cool, neutral hydrogen gas in the ISM/CGM is typically traced by either \textrm{H}\textsc{i} or low-ionization interstellar (LIS) absorption features, while warm/hot ionized gas is probed by high-ionization absorption features. In addition, the strength of the LIS lines provides key information on the properties of the ISM/CGM. Unsaturated lines can be used to infer the ionic column density while saturated lines are useful for probing outflow kinematics and estimating metal covering fractions. In terms of kinematics, due to ubiquitously observed outflows in intensely star-forming galaxies, especially at $z\gtrsim2$, LIS absorption lines tracing foreground gas are found to be blueshifted \citep{Pettini2001,Steidel2010,Du2018}. 
The Ly$\alpha$ feature has a more complex nature. Its morphology ranges from absorption to emission \citep{Shapley2003,Kornei2010}, and the overall strength is determined by the intrinsic production of Ly$\alpha$ photons, and the transfer through both \textrm{H}\textsc{i} gas and dust in the ISM and CGM \citep[e.g.,][]{Trainor2019}.

Empirically, Ly$\alpha$ and LIS equivalent widths (EW$_{Ly\alpha}$ and EW$_{LIS}$, respectively), along with dust extinction, have been used to inform the structure of the CGM. Ly$\alpha$ photons originate from recombination in the \textrm{H}\textsc{ii} regions ionized by massive stars. As Ly$\alpha$ photons propagate through ISM/CGM, they are resonantly scattered by the intervening \textrm{H}\textsc{i} gas and absorbed by dust along their paths. Therefore, the emergent EW$_{Ly\alpha}$ and escape fraction place critical constraints on the covering fraction of the \textrm{H}\textsc{i} gas and the distribution of dust. By examining the strength and scatter of the mutual correlations among EW$_{Ly\alpha}$, EW$_{LIS}$, and $E(B-V)$, we can evaluate the relative importance of different physical processes that shape those relations and infer the key properties of the ISM/CGM in star-forming galaxies.

Metals are believed to be distributed within neutral hydrogen gas, as they are released mainly from supernova explosions. Observationally, the covering fraction of low-ions (or metals that give rise to the LIS transitions) is found to be positively correlated with that of the \textrm{H}\textsc{i} gas \citep{Reddy2016,Gazagnes2018}. This result signals a direct connections between metals and the \textrm{H}\textsc{i} gas, which can naturally be explained if metal-enriched ``pockets" are embedded in the \textrm{H}\textsc{i} gas. It is not yet clear, however, where dust resides in the CGM with respect to the \textrm{H}\textsc{i} gas and metals. Multiple CGM models have also been proposed and tested, which have different assumptions for the geometry and motion of the \textrm{H}\textsc{i} gas (``shell" vs. ``holes"), distribution of dust (uniform screen vs. in gas clumps), and the column density ratio of difference phases of CGM \citep{Neufeld1991,Laursen2013,Duval2014,Gazagnes2018,Steidel2018}.

Correlations involving EW$_{Ly\alpha}$, EW$_{LIS}$, and $E(B-V)$ have been examined both in the nearby universe \citep[e.g.,][]{Rivera-Thorsen2015,Yang2017,Jaskot2019} and at high redshift ($z\gtrsim2$) using individual and composite galaxy spectra \citep{Shapley2001,Shapley2003,Pentericci2007,Pentericci2009,erb2010,Berry2012,Jones2012,Hagen2014,Trainor2016,Du2018,Marchi2019,Trainor2019,Pahl2020}. These previous studies have found that stronger Ly$\alpha$ emission is typically associated with weaker LIS lines and lower dust extinction. The co-dependence on Ly$\alpha$ further suggests a positive correlation between EW$_{LIS}$ and $E(B-V)$. Moreover, while EW$_{Ly\alpha}$ vs. EW$_{LIS}$ and EW$_{Ly\alpha}$ vs. $E(B-V)$ correlations appear to be redshift independent across $z\sim2-4$ \citep{Du2018}, EW$_{Ly\alpha}$ is larger at fixed EW$_{LIS}$ and $E(B-V)$ at $z\sim5$ \citep{Pahl2020}, suggesting greater intrinsic Ly$\alpha$ production at fixed ISM/CGM properties. On the other hand, the 
EW$_{LIS}$ vs. $E(B-V)$ relation does not evolve across $z\sim2-5$ \citep{Pahl2020}, highlighting a fundamental connection between the neutral hydrogen gas and dust components in the ISM/CGM of star-forming galaxies.

The relative scatters or strengths of the correlations among EW$_{Ly\alpha}$, EW$_{LIS}$, and $E(B-V)$ have also been investigated. Using composite spectra created out of nearly 1000 LBGs at $z\sim3$, \citet{Shapley2003} reported that EW$_{Ly\alpha}$ depends more strongly on EW$_{LIS}$ than on $E(B-V)$, as suggested by a larger fractional change in EW$_{LIS}$ across the EW$_{Ly\alpha}$ quartiles than the $E(B-V)$ quartiles. 
More recently, \citet{Du2018} attempted to measure the relative scatter of these 3 correlations in a statistical manner. Treating composites binned according to different properties (i.e., EW$_{Ly\alpha}$, UV absolute magnitude, stellar mass, star-formation rate (SFR), $E(B-V)$, and galaxy age) as ``individual" points, the authors found that EW$_{LIS}$ and EW$_{Ly\alpha}$ are the most strongly correlated.
However, the most robust method for determining the relative degree of intrinsic scatter in these relationships is based on {\it individual} measurements. Such an approach requires significantly higher signal-to-noise for individual rest-frame UV spectra, compared to analyses based on composite spectra.

For example, \citet{Trainor2019} conducted an empirical analysis of factors affecting
Ly$\alpha$ production and escape using a statistical sample of 703 galaxies at $z\sim2-3$. By characterizing and comparing the strengths of various correlations involving EW$_{Ly\alpha}$, EW$_{LIS}$, galaxy properties, the ionization parameter, Ly$\alpha$ kinematics, and Ly$\alpha$ escape fraction, these authors find that EW$_{Ly\alpha}$ is best predicted by a linear combination of EW$_{LIS}$ and [\textrm{O}\textsc{iii}]/H$\beta$. However, since \citet{Trainor2019} primarily focused on the question of 
Ly$\alpha$ production and escape, they did not directly analyze the EW$_{LIS}$ vs. $E(B-V)$ relation, which can in fact provide additional insights into the ISM/CGM structure, as we explore in this work.

In this study we present the first statistical analysis to characterize the nature and relative strength of the mutual correlations involving EW$_{Ly\alpha}$, EW$_{LIS}$, and $E(B-V)$ using rest-UV spectroscopic measurements of individual galaxies at $z\sim2$. We further utilize available rest-optical spectra for the same galaxies to obtain supplementary measurements, such as the systemic redshift and oxygen abundance inferred from rest-optical nebular emission lines, to build a multidimensional description of each galaxy. In order to perform meaningful comparisons among the correlations, we measure EW$_{Ly\alpha}$, EW$_{LIS}$, and $E(B-V)$ for all the objects in a uniform manner. We additionally 
account for individual LIS non-detections to avoid potential selection biases. With a carefully constructed sample and uniform measurements, we aim to determine the most fundamental correlation among the three, and infer the CGM structure and dust distribution based on our results. 

This paper is organized as follows. In Section~\ref{sec:data}, we discuss the sample selection, rest-optical and rest-UV observations, redshift estimates, and the sample properties. We describe the measurements in Section~\ref{sec:measure}, including SED modeling and the EW measurements of Ly$\alpha$ and LIS features. We present the EW$_{Ly\alpha}$ and $E(B-V)$ distributions in Section \ref{sec:results}, along with the statistical analysis and linear regression modeling to the EW$_{Ly\alpha}$ vs. EW$_{LIS}$, EW$_{LIS}$ vs. $E(B-V)$, and EW$_{Ly\alpha}$ vs. $E(B-V)$ correlations. In Section~\ref{sec:discussion}, we discuss plausible CGM models suggested by our results and the factors contributing to the observed scatter in the correlations. We summarize the key findings in Section~\ref{sec:summary}.

Throughout this paper, we adopt a standard $\Lambda$CDM model
with $\Omega_{m}=0.3$, $\Omega_{\Lambda}=0.7$, and $H_{0}=70$ \kms, and a solar oxygen abundance of 12+log(O/H)=8.69 \citep{Asplund2009}. All wavelengths are measured in vacuum. Magnitudes and colors are on the AB system.

\section{Sample and Observations}
\label{sec:data}

In this section, we describe the parent MOSDEF-LRIS dataset, the systemic redshift determination, and the properties of the LRIS-ISM sample used in this study. 
For a more in-depth description of sample selection, data reduction, and redshift measurements, we refer readers to \citet{Kriek2015} and \citet{Topping2020}, in which the rest-optical and rest-UV data, respectively, were first presented.

\subsection{Data and Observations}
\label{sec:obs}

\subsubsection{Rest-Optical MOSFIRE Spectroscopy}
\label{sec:mosfire}

The MOSDEF-LRIS sample presented in this paper was drawn from the MOSFIRE Deep Evolution Field (MOSDEF) survey described in \citet{Kriek2015}, which consists of $\sim 1500$ galaxies at $1.4 \leq z \leq 3.8$. The targets were observed on 53 multi-object slitmasks using the Multi-object Spectrometer for Infrared Exploration \citep[MOSFIRE;][]{McLean2012} on the Keck I telescope between 2012 to 2016. Rest-frame optical spectra were obtained with $Y$, $J$, $H$, and $K$ filters, which optimize the coverage of strong rest-optical emission lines ([\textrm{O}\textsc{ii}]$\lambda\lambda$3727,3729, H$\beta$, [\textrm{O}\textsc{iii}]$\lambda\lambda4959,5007$, H$\alpha$, [\textrm{N}\textsc{ii}]$\lambda6584$, and [\textrm{S}\textsc{ii}]$\lambda\lambda6717,6731$) for objects in redshift windows $1.37\leqslant z \leqslant 1.70$, $2.09\leqslant z \leqslant 2.61$, and $2.95\leqslant z \leqslant 3.80$.

The targets were $H$-band selected and have an approximate  corresponding lower stellar mass limit of $\sim10^{9}M_{\sun}$ in each redshift bin. The MOSDEF targets are located in extragalactic legacy fields covered by the CANDELS \citep{Grogin2011,Koekemoer2011} and 3D-HST surveys \citep{Brammer2012}, where extensive spectroscopic and multiwavelength photometric data are publicly available and enable a multi-dimensional view of each galaxy. The MOSFIRE spectra were reduced, optimally extracted, and placed on an absolute flux scale as described in \citet{Kriek2015}.

\subsubsection{Rest-UV LRIS Spectroscopy}
\label{sec:LRIS_spec}

A detailed description of the rest-UV sample selection, data collection, and data reduction is provided in \citet{Topping2020}, and here we only provide an overview. In summary, a subset of MOSDEF galaxies were selected for  rest-UV spectroscopic followup using the Low Resolution Imager and Spectrometer \citep[LRIS,][]{Oke1995,Steidel2004} on the Keck I telescope. The followup observations prioritized targets with individually detected strong rest-optical lines (H$\beta$, [\textrm{O}\textsc{iii}], H$\alpha$, and [\textrm{N}\textsc{ii}]). We also included MOSDEF galaxies with confirmed spectroscopic redshifts and objects from the 3D-HST catalog with similar photometric redshifts and apparent magnitudes to those of the MOSDEF galaxies. A total of 260 galaxies were observed within the redshift ranges of $1.40\leqslant z \leqslant 1.90$, $1.90\leqslant z \leqslant 2.65$, and $2.95\leqslant z \leqslant 3.80$. The vast majority of MOSDEF-LRIS galaxies (215 out of 260) have spectroscopic redshift measurements inferred from rest-optical nebular emission lines in the MOSFIRE spectra. Of the remaining 45 objects (hereafter the ``LRIS-only" galaxies), 21 have a redshift estimated based on rest-UV emission and absorption features (see Section \ref{sec:z_meas}). The remaining 24 ``LRIS-only" galaxies do not have robust redshift measurements.

Multi-slit rest-UV spectroscopy was obtained using Keck/LRIS, a dichroic spectrograph, in the COSMOS, AEGIS, GOODS-N, and GOODS-S fields, over the course of 10 nights in 2017 and 2018. LRIS data were collected on 9 multi-object slitmasks with $1.''2$ slits. 
All masks were observed with the 400 lines $\mbox{mm}^{-1}$ grism blazed at 3400$\mbox{\AA}$ on the blue side (435 $\mbox{km s}^{-1}$ FWHM), and the 600 lines $\mbox{mm}^{-1}$ grating blazed at 5000$\mbox{\AA}$ on the red side (220 $\mbox{km s}^{-1}$ FWHM). This configuration enabled continuous wavelength coverage from 3100$\mbox{\AA}$ to $\sim$7650$\mbox{\AA}$ in the observed frame, covering a number of strong rest-UV spectral lines given the redshift ranges probed. These include Ly$\alpha$,  low-ionization \textrm{Si}\textsc{ii}$\lambda1260$,  \textrm{O}\textsc{i}$\lambda1302$+\textrm{Si}\textsc{ii}$\lambda1304$,  \textrm{C}\textsc{ii}$\lambda1334$, \textrm{Si}\textsc{ii}$\lambda1526$,  \textrm{Fe}\textsc{ii}$\lambda1608$, and \textrm{Al}\textsc{ii}$\lambda1670$ absorption features, high-ionization \textrm{Si}\textsc{iv}$\lambda\lambda$1393,1402 and \textrm{C}\textsc{iv}$\lambda\lambda$1548,1550 absorption features, and \textrm{O}\textsc{iii}]$\lambda\lambda$1661,1665,  \textrm{He}\textsc{ii}$\lambda$1640, and \textrm{C}\textsc{iii}]$\lambda\lambda$1907,1909 emission features. The integration time ranged from 6 to 11 hours for different masks, with a median of $\sim7.5$ hours. Observing conditions were fair, yielding a moderate (average $0.''8$) seeing.

The data were reduced using customized {\it IRAF}, {\it IDL}, and {\it Python} scripts as described in \citet{Topping2020}. In brief, the two-dimensional (2D) spectra were rectified, flatfielded, cut up into individual slitlets, cleaned of cosmic rays, background-subtracted, and median stacked. The relative order of the last 3 steps listed above was slightly different for the blue- and red-side slitlets to optimize the data quality, given that the red side is more heavily affected by cosmic rays \citep[see][for details]{Topping2020}. To avoid the overestimation of the background due to the presence of the target \citep{Shapley2006}, a second-pass background subtraction was performed for each object, during which the trace  determined from the stacked 2D spectrum was masked out. Afterwards, the 2D spectra were extracted into one dimension (1D), and wavelength and flux calibrated. The flux calibration included two steps, a relative calibration using a spectrophotometric standard star, and an absolute calibration to scale our spectrophotometric measurements to those listed in the 3D-HST  photometric catalog. Finally, additional continuum correction was applied to a small number of objects to ensure the consistency of the continuum levels on either side of the dichroic at $\sim$5000$\mbox{\AA}$. 

\subsection{Redshift Measurements}
\label{sec:z_meas}

For most (215 out of 260) objects in the MOSDEF-LRIS sample, the systemic redshift was robustly measured using the rest-optical emission lines (H$\beta$, [\textrm{O}\textsc{iii}], H$\alpha$, [\textrm{N}\textsc{ii}], and [\textrm{S}\textsc{ii}]), with the initial guess derived from the highest S/N emission line, typically being H$\alpha$ or [$\textrm{O}\textsc{iii}]\lambda5007$ \citep{Kriek2015}. For 21 out of 45 LRIS-only objects, with no previous MOSFIRE redshift, a systemic redshift measurement was obtained based on Ly$\alpha$ emission and/or LIS absorption lines (\textrm{Si}\textsc{ii}$\lambda1260$,  \textrm{O}\textsc{i}$\lambda1302$+\textrm{Si}\textsc{ii}$\lambda1304$,  \textrm{C}\textsc{ii}$\lambda1334$, \textrm{Si}\textsc{ii}$\lambda1526$,  \textrm{Fe}\textsc{ii}$\lambda1608$, and \textrm{Al}\textsc{ii}$\lambda1670$), as described in detail in \citet{Topping2020}. A variety of velocity rules were applied to the Ly$\alpha$- and LIS-based redshifts ($z_{Ly\alpha}$ and $z_{LIS}$, respectively) in order to account for the presence of galaxy-scale outflows. $z_{LIS}$ was used to calculate the systemic redshift whenever available, and assumed to have a blueshift of $-32$ \kms and $-89$ \kms, respectively, for objects with only $z_{LIS}$ (8 galaxies) and with both $z_{LIS}$ and $z_{Ly\alpha}$ measurements (5 galaxies). In cases where only $z_{Ly\alpha}$ was available, Ly$\alpha$ was assumed to have a redshift of $+153$ \kms and $+317$ \kms for objects at $z\leqslant2.7$ (7 galaxies) and $z>2.7$ (1 galaxy), respectively. All the observed-frame spectra were then shifted into the rest frame according to the systemic redshift determined using the methods above.

\subsection{Sample}
\label{sec:sample}

\begin{figure*}
\includegraphics[width=1.0\linewidth]{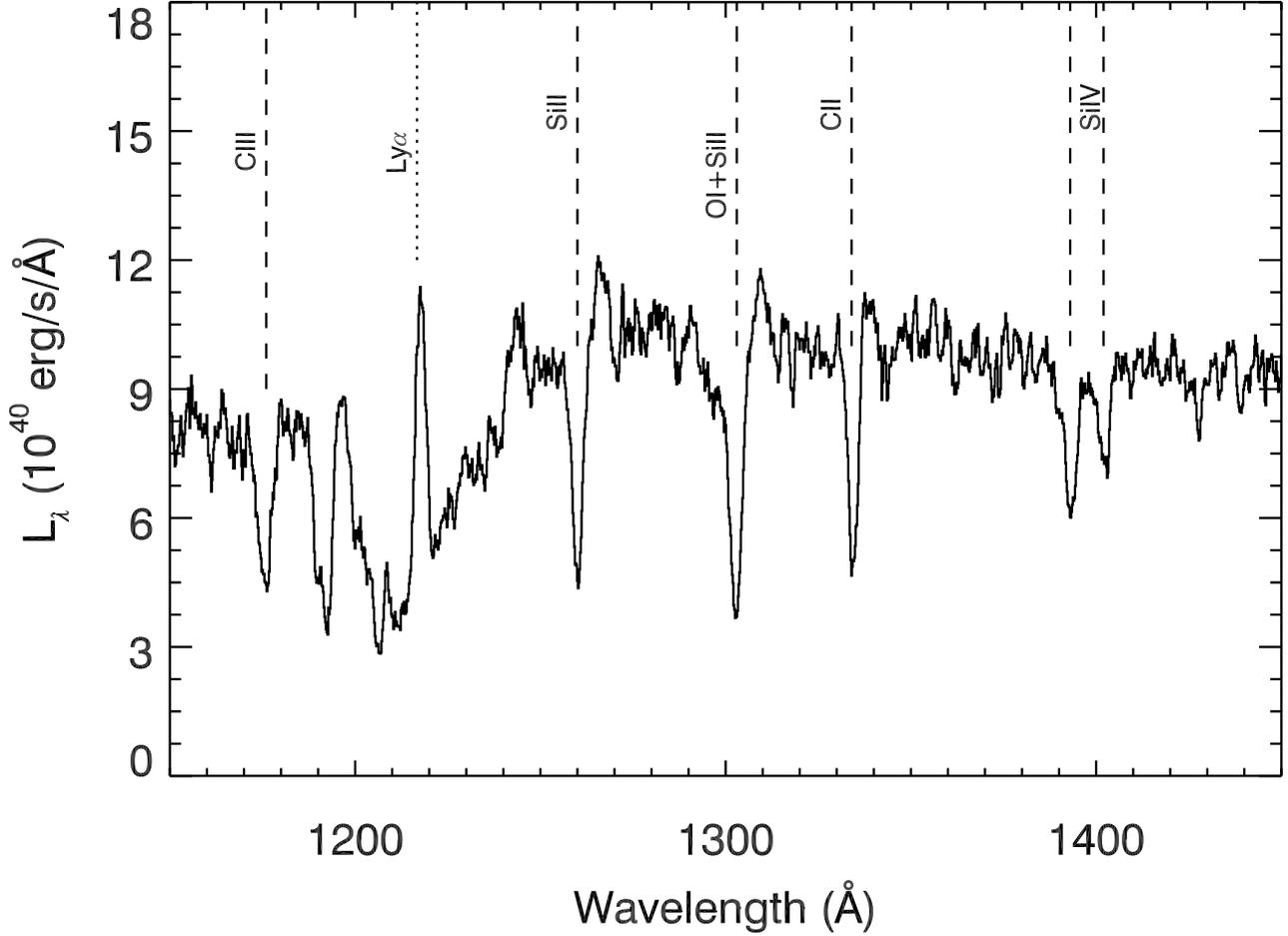}
\caption{Composite rest-frame far-UV spectrum of the 157 galaxies in the LRIS-ISM sample. We show the rest-frame $1150\mbox{\AA}-1450\mbox{\AA}$ portion of the composite spectrum, as only the spectral region from Ly$\alpha$ to \textrm{C}\textsc{ii}$\lambda1334$ has contribution from all 157 galaxies. Key emission (dotted lines) and absorption (dashed lines) features in this range are labeled.}
\label{fig:spec}
\end{figure*}

\begin{figure*}
\includegraphics[width=0.5\linewidth]{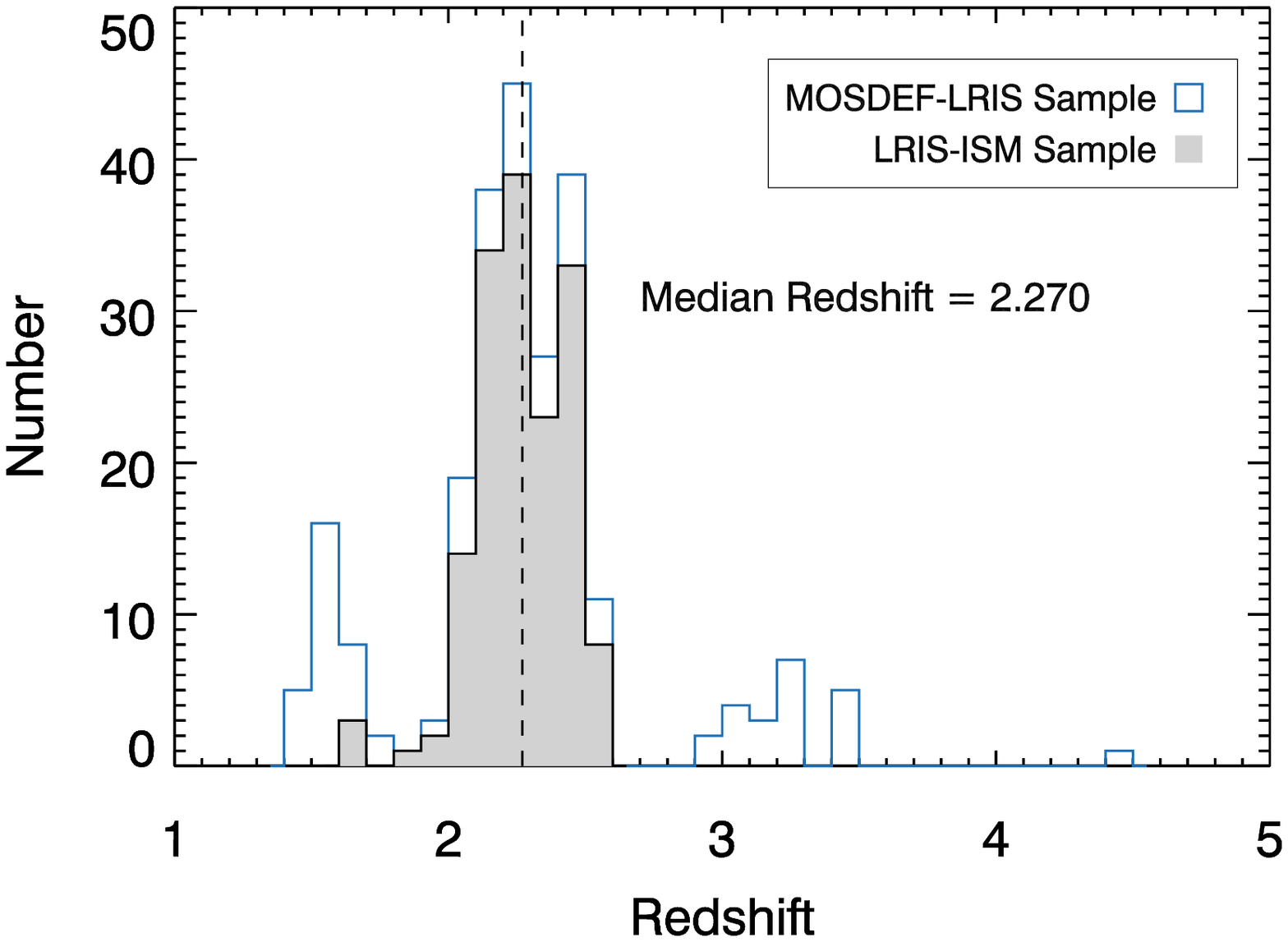}
\includegraphics[width=0.5\linewidth]{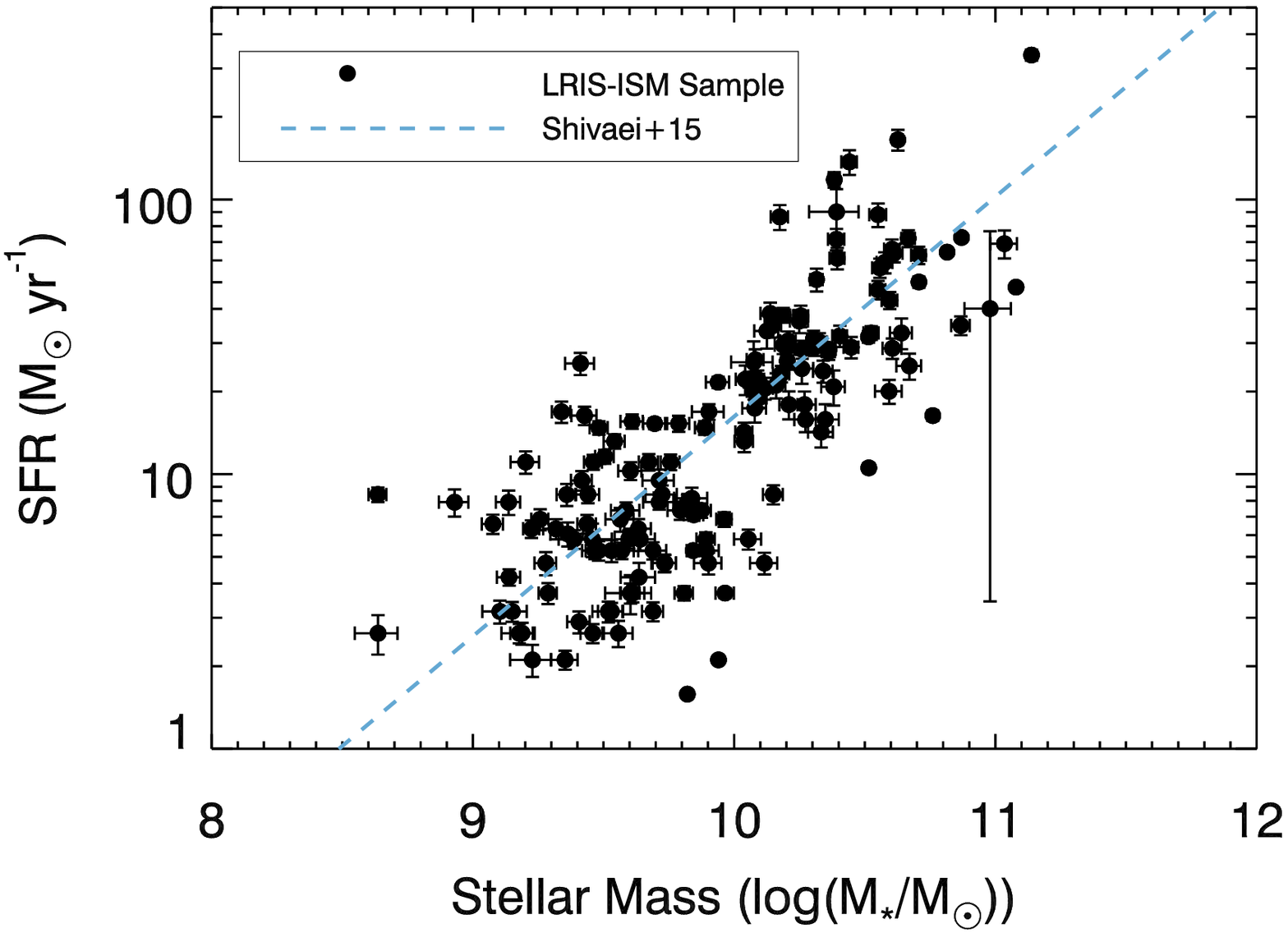}
\caption{Galaxy properties of the LRIS-ISM sample. \textbf{Left:} Redshift distribution. Open blue and solid gray bars represent the parent MOSDEF-LRIS sample with redshift measurements (either from MOSFIRE or LRIS; 236 objects) and the LRIS-ISM sample (157 objects), respectively. The vertical dashed line marks the median redshift of the LRIS-ISM sample. \textbf{Right:} SFR vs. stellar mass. Both SFR and stellar mass are derived from SED modeling assuming a Chabrier IMF, as described in Section \ref{sec:sed}. Error bars in SFR and stellar mass indicate the associated 1$\sigma$ uncertainties. For comparison, we indicate with a blue dashed line the SFR vs. stellar mass relation derived in \citet{Shivaei2015} using objects with $M_{*}>10^{9.5}M_{\sun}$ in the parent MOSDEF sample. While \citet{Shivaei2015} presented multiple SFR estimates (e.g., UV-, H$\alpha$-, and SED-based), we only plot those based on SED modeling for a direct comparison with our work.} 
\label{fig:galprop}
\end{figure*}

In this study, we aim to examine the mutual correlations among EW$_{Ly\alpha}$, EW$_{LIS}$, and dust reddening $E(B-V)$. For this purpose, we selected a subsample of galaxies that have simultaneous coverage of Ly$\alpha$, \textrm{Si}\textsc{ii}$\lambda1260$, \textrm{O}\textsc{i}$\lambda1302$+\textrm{Si}\textsc{ii}$\lambda1304$, and \textrm{C}\textsc{ii}$\lambda1334$, which corresponds to the rest-frame wavelength range of $1208-1344\mbox{\AA}$. In order to retain a sufficiently large sample, we did not further require the coverage of the redder LIS lines, namely, \textrm{Si}\textsc{ii}$\lambda1526$, \textrm{Fe}\textsc{ii}$\lambda1608$, and \textrm{Al}\textsc{ii}$\lambda1670$. Calculating EW$_{LIS}$ based on \textrm{Si}\textsc{ii}$\lambda1260$, \textrm{O}\textsc{i}+\textrm{Si}\textsc{ii}, and \textrm{C}\textsc{ii} additionally enables a direct comparison with the results presented in \citet{Du2018}, where the relations among EW$_{Ly\alpha}$, EW$_{LIS}$, and $E(B-V)$ were examined in stacked spectra with the LIS EW measured from the same 3 features. The spectral coverage requirement excludes 15 objects from the sample.
Moreover, to avoid potential redshift evolution in the relations of interest, we only included objects with a systemic redshift $z\leqslant2.7$, which is consistent with the redshift boundary chosen for the $z\sim2$ sample presented in \citet{Du2018}. With this redshift threshold, all the relevant lines (Ly$\alpha$ and 3 LIS lines) fall on the blue-side of the spectrum. Twenty-two $z>2.7$ galaxies were removed from the sample according to this criterion. We also imposed a few more criteria to exclude objects for which the spectra are subject to nearby contamination and/or artifacts (27 galaxies) and those identified as AGNs (13 galaxies) based on X-ray luminosities, $Spitzer/IRAC$ colors, or [\textrm{N}\textsc{ii}]$\lambda$6584/H$\alpha$ ratios \citep{Coil2015,Azadi2017,Azadi2018,Leung2019}. Finally, we removed two objects with anomalously large error bars on their EW$_{Ly\alpha}$ measurements: COSMOS-4029 (EW$_{Ly\alpha}$=$21.9\pm296.0\mbox{\AA}$) and GOODS-N-29834 (EW$_{Ly\alpha}$=$195.3\pm1125.0\mbox{\AA}$). These two galaxies both have a close-to-zero red-side continuum ($1225-1255\mbox{\AA}$; see Section \ref{sec:lya} for details of Ly$\alpha$ EW measurement) and a noise level comparable to the continuum. Therefore, the exclusion of these two objects from the final sample was based on the fact that we were unable to obtain meaningful measurements of EW$_{Ly\alpha}$ for them.

In the end, the selected subsample (hereafter the ``LRIS-ISM" sample) consists of 157 galaxies,  of which 16 are LRIS-only objects. 
We summarize the galaxy properties and line measurements of the LRIS-ISM objects in Table \ref{tab:galprop} and plot the overall rest-UV composite spectrum in the wavelength range of interest in Figure \ref{fig:spec}. As shown in Figure \ref{fig:galprop} and the lower panel of Figure \ref{fig:ew_hist}, our LRIS-ISM sample has a redshift range of $1.65 \leq z \leq 2.58$ with a median of 2.27, a stellar mass range of $8.64<$ log$(M_{*}/M_{\sun})<11.14$ with a median of 9.94, a star formation rate (SFR) range of from $2 \leq (\mbox{SFR}/M_{\sun}\mbox{ yr}^{-1}) \leq 336$ with a median of 15 M$_{\sun}$yr$^{-1}$, and dust reddening range of 0.01 $< E(B-V) <$ 0.435 with a median of 0.10. We describe the SED modeling in Section \ref{sec:sed}.

\section{Measurements}
\label{sec:measure}

One unique aspect of our study is the ability to characterize the relative tightness of the relations among
EW$_{Ly\alpha}$, EW$_{LIS}$, and $E(B-V)$ using measurements from \textit{individual} galaxies, in order to gain insights into the distribution of gas and dust in distant star-forming galaxies. In this section, we describe the spectral energy distribution (SED) modeling for obtaining dust reddening and other stellar population parameters (Section \ref{sec:sed}), as well as the measurements of rest-frame Ly$\alpha$ EW (Section \ref{sec:lya}) and the total EW of multiple LIS features (Section \ref{sec:lis}).

\subsection{SED Modeling}
\label{sec:sed}

We inferred the key parameters (such as dust reddening, stellar mass, age, and SFR) of individual galaxies in the LRIS-ISM sample by modeling their broadband photometry, as cataloged  by the 3D-HST survey \citep{Skelton2014}. When applicable, the photometry was corrected for rest-optical
([\textrm{O}\textsc{ii}], H$\beta$, [\textrm{O}\textsc{iii}], H$\alpha$, and [\textrm{N}\textsc{ii}]) and Ly$\alpha$ line emission prior to the fitting.

Galaxy SEDs were fit with stellar population templates from \citet {BC03} assuming a \citet{Chabrier2003} initial mass function (IMF). Following \citet{Reddy2017,Reddy2018a} and \citet{Du2018}, we modeled each galaxy with two different combinations of metallicity and extinction curves: 1.4 solar metallicity ($Z_{\sun}=0.014$) with the \citet{Calzetti2000} attenuation curve (hereafter ``1.4 $Z_{\sun}$+Calzetti"), and 0.28 $Z_{\sun}$ with the SMC extinction curve (hereafter ``0.28 $Z_{\sun}$+SMC"). Although each model grid included different star-formation histories (exponentially declining, constant, and rising) and wide ranges in age and dust reddening, we chose to adopt the constant SFR model, as it is proven to be a satisfactory description of the star-formation history for the typical star-forming galaxies at $z\sim2$ \citep{Reddy2012,Steidel2014,Strom2017}. We set a lower age limit of 50 Myr, provided the typical dynamical timescales of $z\sim2$ galaxies \citep{Reddy2012}, and an upper age limit of the age of the universe at the redshift of each galaxy.
We also considered stellar continuum reddening in the range $0.0<E(B-V)<0.6$.

Recent work suggests that sub-solar metallicity models with an SMC curve provide a better description for low- and moderate-mass star-forming galaxies at $z\gtrsim2$ than the traditionally assumed combination of solar metallicity and a \citet{Calzetti2000} curve, based on the IRX-$\beta$ relation \citep[e.g.,][]{Reddy2017,Reddy2018a} and the overall fit to the galaxy SEDs \citep{Du2018}. As a result, we adopted the 0.28 $Z_{\sun}$+SMC model for the LRIS-ISM galaxies with log$(M_{*}/M_{\sun})<10.04$ and the 1.4 $Z_{\sun}$+Calzetti model for those with log$(M_{*}/M_{\sun})\geqslant10.04$, where log$(M_{*}/M_{\sun})$ was estimated based on the 1.4 $Z_{\sun}$+Calzetti model. This stellar-mass threshold corresponds to nebular dust reddening $E(B-V)_{neb}=0.3$, according to the $E(B-V)_{neb}$ vs. $M_{*}$ relation in \citet{Shivaei2020}, and we estimated $E(B-V)_{neb}$ for each galaxy based on the Balmer decrement assuming a \citet{Cardelli1989} law. 

To obtain the best-fit stellar population parameters and their uncertainties, we created 100 realizations of the SED for each galaxy by perturbing its photometric measurements with the associated 1$\sigma$ errors listed in the 3D-HST photometric catalog. The same modeling described above was performed on these 100 realizations, resulting in 100 sets of best-fit parameters for each object determined by the minimum $\chi^2$ method. The median value and standard deviation of those 100 measurements were adopted as the best-fit stellar population parameter and the 1$\sigma$ uncertainty, respectively.

\subsection{Line Measurements}
\label{sec:line}

\subsubsection{Ly$\alpha$}
\label{sec:lya}

We measured the rest-frame $\mbox{Ly}\alpha$ EW following the procedures described in \citet{Kornei2010} and \citet{Du2018}. The spectral morphology of $\mbox{Ly}\alpha$ in individual galaxy spectra was classified into 4 categories through visual inspection: ``emission," ``absorption," ``combination," and ``noise."  ``Emission" objects show dominant Ly$\alpha$ emission on top of a relatively flat continuum, while the Ly$\alpha$ emission for ``combination" objects is superimposed on a large absorption trough. The Ly$\alpha$ morphology is classified as ``absorption" when a broad absorption trough resides around the rest-frame wavelength of Ly$\alpha$, and as ``noise" when the spectrum is featureless near Ly$\alpha$. 

Galaxies in the LRIS-ISM sample were mainly categorized as ``combination" (61 galaxies), ``absorption" (48 galaxies) and ``noise" objects (38 galaxies), with only a small fraction identified as ``emission" objects (10 galaxies). For each object, regardless of their spectral morphology, the blue and red side continuum levels were estimated over the wavelength range of $1120-1180\mbox{\AA}$ and $1225-1255\mbox{\AA}$, respectively. For ``emission,"``combination," and ``absorption" objects, the $\mbox{Ly}\alpha$ flux was integrated between the blue and red wavelength ``boundaries," where the flux density level on either side of the $\mbox{Ly}\alpha$ feature (either emission or absorption) first meets the blue and red side continuum level, respectively. The blue boundary was fixed at $1208\mbox{\AA}$ for the ``emission" objects and forced to be no bluer than $1208\mbox{\AA}$ for the ``combination" objects \citep{Du2018}. For ``noise" objects, the Ly$\alpha$ flux was integrated over 1199.9 to 1228.8 $\mbox{\AA}$, the boundaries adopted in \citet{Kornei2010}. Finally, we computed $\mbox{Ly}\alpha$ EW by dividing the enclosed $\mbox{Ly}\alpha$ flux by the red side continuum flux-density level. Although by construction, all galaxies in the LRIS-ISM sample have Ly$\alpha$ coverage, 4 galaxies (1 ``emission" and 3 ``noise" objects) have no spectral coverage at rest-frame wavelengths shorter than $1160\mbox{\AA}$, such that the spectral region available is not sufficient (less than 20$\mbox{\AA}$) for a robust estimate of the blue-side continuum level. Consequently, we measured, in each Ly$\alpha$ morphology category, the relative level of the blue and red side continua from objects with adequate spectral coverage on both sides. The median blue-to-red continuum ratio in corresponding Ly$\alpha$ morphology category was then applied as a rough proxy of the blue continuum for those 4 objects lacking sufficient blue-side spectral coverage. 

To characterize the uncertainty on Ly$\alpha$ EW, we perturbed the science spectrum 100 times with its corresponding error spectrum, and measured the Ly$\alpha$ EW in the 100 fake spectra for each galaxy. The sigma-clipped average and standard deviation of those 100 measurements were adopted as the final Ly$\alpha$ EW and the $1\sigma$ uncertainty, respectively, for each galaxy. We list the measured rest-frame Ly$\alpha$ EW for all 157 galaxies in the LRIS-ISM sample in Table \ref{tab:galprop}.

\subsubsection{LIS lines}
\label{sec:lis}

Our deep LRIS spectra enable LIS line measurements from individual objects in the LRIS-ISM sample. In this study we focus on the EW of the 3 LIS lines near Ly$\alpha$: \textrm{Si}\textsc{ii}$\lambda$1260, \textrm{O}\textsc{i}$\lambda$1302+\textrm{Si}\textsc{ii}$\lambda$1304, and \textrm{C}\textsc{ii}$\lambda$1334. These lines are among the strongest rest-UV LIS features and typically saturated in the MOSDEF-LRIS sample. Provided that we require measurements of EW$_{Ly\alpha}$ and EW$_{LIS}$, measuring these 3 lines would maximize the sample size because of their proximity to the Ly$\alpha$ feature. Given the low resolution of the blue-side LRIS spectra, \textrm{O}\textsc{i}$\lambda$1302+\textrm{Si}\textsc{ii}$\lambda$1304 is blended in the individual spectra. As in \citet{Du2016}, we adopted single-component Gaussian fits as the simplest but sufficient functional form to describe the interstellar absorption line profiles, because the data are not of sufficient resolution to consider more complex models. We have further tested that for obtaining EW measurements, single-component Gaussian fits yield almost identical results to those suggested by two-component fits (i.e., two independent Gaussians describing the systemic and outflowing components, respectively).

We continuum normalized the rest-frame composite spectra using spectral windows that are clean of spectral features defined by \citet{Rix2004}. Based on these windows, we modeled the continuum for all composite spectra with the IRAF $continuum$ routine, using a $spline\it3$ function of order $=8$. Regions near Ly$\alpha$ (1197$\mbox{\AA}$ to 1248 $\mbox{\AA}$)) were not used for estimating the continuum, as the continuum is fairly curved near that region and affected by the broad Ly$\alpha$ absorption trough for the ``absorption" and ``combination" objects. In cases where the fitted continuum level did not provide a proper description of the observed spectrum due to the limited coverage of windows from \citet{Rix2004}, additional windows customized for each object were added to provide reasonable constraints on the fit.

The absorption line profile fitting was performed on the continuum-normalized composite spectra. For the ``detection," ``partial detection," and ``combined detection" objects (see descriptions below), we used the IDL program MPFIT \citep{Mark2009} with the initial values 
of continuum flux level, line centroid, EW and Gaussian FWHM estimated from the program $splot$ in IRAF. The best-fit was then determined where the $\chi^{2}$ of the fit reached a minimum. We iterated the fitting over a narrower wavelength range for all the interstellar absorption lines: centroid$-4\sigma < \lambda < $centroid$+4\sigma$, where the centroid and $\sigma$ are, respectively, the returned central wavelength and standard deviation of the best-fit Gaussian profile from the initial MPFIT fit to respective lines over $\lambda_{rest}-10\mbox{\AA}$ to $\lambda_{rest}-10\mbox{\AA}$. 

Ideally, we would like to measure \textrm{Si}\textsc{ii}, \textrm{O}\textsc{i}+\textrm{Si}\textsc{ii}, and \textrm{C}\textsc{ii} individually and take the sum of their EWs as EW$_{LIS}$ for each object. However, not all 3 lines are detected at the $\geqslant3\sigma$ level for every galaxy in the LRIS-ISM sample. Therefore, we inferred the total LIS EW differently for objects in four different LIS categories: 

$Detection$: 51 out of 157 objects, marked as `D' in Table \ref{tab:galprop} and the figures in Section \ref{sec:results}. In the spectra for these objects, all 3 LIS lines are individually detected. Therefore, the total EW of the LIS lines was adopted as EW$_{LIS}$ for such objects, and the $1\sigma$ uncertainty was estimated by adding the $1\sigma$ error bar of all 3 LIS lines in quadrature. 

$Partial$ $Detection$: 61 out of 157 objects, marked as `P'. In the spectra for these objects,  1 or 2 LIS lines are individually detected, even though we required all 3 LIS lines to be covered in individual galaxy spectra. This is because we excluded particular LIS lines, in a small number of cases, when they were clearly contaminated or showed an unphysical absorption profile (e.g., significantly negative at maximum depth). Given that the LIS lines are typically saturated in star-forming galaxies with properties typical of those in our sample \citep[e.g.,][]{Shapley2003}, their EWs are not sensitive to column density but instead the combination of the covering fraction and velocity dispersion of the respective ions. To that end, we expect the lines we measure here (\textrm{Si}\textsc{ii}, \textrm{O}\textsc{i}+\textrm{Si}\textsc{ii}, and \textrm{C}\textsc{ii}) have the same relative strengths across this category, assuming all low ions have similar covering fractions and velocity dispersions. For ``partially detected" objects, the undetected LIS lines primarily result from systematic errors or artifacts in the spectra but not their weaker nature. To infer the total EW$_{LIS}$, we calculated $f_{i}$, the fractional contribution of the EW of absorption line $i$ to EW$_{LIS}$ in the sample of ``detection" objects. Explicitly, $f_{i}=EW_{i}/EW_{LIS}$, where $i$ denotes respective LIS lines and $EW_{i}$ is the measured EW of that line. The median $f$ values and associated uncertainties, $\Delta$$f$, 
for \textrm{Si}\textsc{ii}, \textrm{O}\textsc{i}+\textrm{Si}\textsc{ii}, and \textrm{C}\textsc{ii} are, respectively, 0.285$\pm$0.063, 0.422$\pm$0.077, and 0.287$\pm$0.056, where the uncertainties were estimated from error propagation. Those median $f$ values were used to scale the measured partial EWs to a ``total" EW$_{LIS}$. For example, in the case of an object with only a detected \textrm{O}\textsc{i}$\lambda1302$+\textrm{Si}\textsc{ii}$\lambda1304$ line, we calculated EW$_{LIS}$ by dividing the \textrm{O}\textsc{i}+\textrm{Si}\textsc{ii} EW by 0.422. Similarly, for an object with detections in \textrm{Si}\textsc{ii}$\lambda1260$ and \textrm{C}\textsc{ii}$\lambda1334$, we calculated its EW$_{LIS}$ by dividing the sum of \textrm{Si}\textsc{ii} and \textrm{C}\textsc{ii} EWs by $0.285+0.287=0.572$. The uncertainty on the inferred total $EW_{LIS}$ for individual objects was obtained through propagation of error using the following equation, determined from the sample of ``detected" objects:

\begin{equation}
\Delta EW_{LIS} = EW_{LIS}\sqrt{(\frac{\Delta f_{i}}{f_{i}})^2+(\frac{\Delta EW_{i}}{EW_{i}})^2}
\end{equation}

where $\Delta$$EW_{i}$ represents the measurement uncertainty of line $i$ (when only 1 line is detected) or the combination of lines (e.g., \textrm{Si}\textsc{ii}$\lambda1260$ and \textrm{C}\textsc{ii}$\lambda1334$; when 2 lines are detected). The fractional contributions of a combination of 2 detected lines to $EW_{LIS}$ are $0.713\pm0.111$, $0.715\pm0.110$, and $0.578\pm0.094$ for \textrm{Si}\textsc{ii}$\lambda1260$+\textrm{O}\textsc{i}$\lambda1302$$+$\textrm{Si}\textsc{ii}$\lambda1304$, \textrm{O}\textsc{i}$\lambda1302$+\textrm{Si}\textsc{ii}$\lambda1304$$+$\textrm{C}\textsc{ii}$\lambda1334$, and \textrm{Si}\textsc{ii}$\lambda1260$$+$\textrm{C}\textsc{ii}$\lambda1334$, respectively.

The above method of determining the overall uncertainly on $EW_{LIS}$ based on partial information accounts for not only the measurement uncertainties from individual LIS lines, but also that associated with respective scaling factors, $f$. As described below, we adopted the same error estimate method for ``combined detection" and ``limit" objects when not all 3 LIS lines are available. We note that the absolute values of LIS measurement uncertainties do not affect the relative strengths of the mutual correlations among EW$_{Ly\alpha}$, EW$_{LIS}$, and $E(B-V)$, as suggested by additional $ASURV$ tests (see Section \ref{sec:corr} for details).

$Combined$ $Detection$: 12 out of 157 objects, marked as `C'. While no LIS lines are individually detected, the combined line $S/N$ (from 2 or 3 lines) is $\geqslant3$. For objects with all 3 lines available, the total EW and the associated $1\sigma$ uncertainty on EW$_{LIS}$ were estimated using the same method as for the ``detection" objects. In cases where only 2 lines were available and showed a combined line $S/N \geqslant3$, we computed its total EW and the associated $1\sigma$ uncertainty by scaling them using the same corresponding factors calculated in the ``partial detection" category (see above).

$Limit$: 33 out of 157 objects, marked as `L'. In this category, objects have no LIS lines individually detected, and the combined line $S/N$ (from 1, 2, or 3 lines) is $<3$. Considering that the non-detected LIS lines may not have a well defined Gaussian profile, we performed simple integration over a spectral range of $\pm 2\mbox{\AA}$ from the rest-frame wavelength of each LIS line on the continuum-normalized spectra. This $4\mbox{\AA}$-wide window has been tested to be sufficient to capture the LIS non-detections, based on the typical width of LIS features measured at the $2-3\sigma$ level in individual continuum-normalized spectra. For objects in the ``limit" category, we report a $3\sigma$ upper limit in EW$_{LIS}$. We note that, in such cases, EW$_{LIS}$ is plotted in figures as a lower limit, since we define LIS absorption as negative. The $1\sigma$ uncertainty on EW$_{LIS}$ was estimated by adding the $1\sigma$ error bar of all available LIS lines in quadrature, where the error bar on individual LIS EW was calculated by adding the continuum-normalized flux density level at each wavelength in quadrature in the corresponding error spectrum over the same spectral region used for the flux integration (i.e., line centroid $\pm 2\mbox{\AA}$). In cases where only 1 or 2 LIS lines are available, we scaled the limit using the same corresponding factors calculated in the ``partial detection" category (see above).

In summary, 71$\%$ of the LRIS-ISM sample has EW$_{LIS}$ inferred from robust detections of one or more lines, while only 21$\%$ has limits and 8$\%$ requires combining LIS lines to yield a detection.  Obtaining robust LIS measurements from firm detections for a large majority of the sample enables us to examine the intrinsic scatter of the relations presented in Section \ref{sec:results} from an individual-object perspective.

\LongTables
\begin{deluxetable*}{ccccccc}
\tablewidth{0pt}
  \tablecaption{Galaxy Properties and Line Measurements}
  \tablehead{
    \colhead{Field} &
    \colhead{ID} &
    \colhead{Redshift} &
    \colhead{$E(B-V)$} &
   \colhead{EW$_{{Ly}\alpha}$} &
    \colhead{EW$_{LIS}$} & 
    \colhead{LIS Category} \\
    \colhead{} &
    \colhead{} &
    \colhead{} &
    \colhead{} &
    \colhead{($\mbox{\AA}$)} &
    \colhead{($\mbox{\AA}$)} &
    \colhead{} \\
    }
  \startdata

AEGIS &      3668 &     2.1877 &      0.110 $\pm$      0.011 &        0.6 $\pm$        0.6 &       -9.7 $\pm$        0.4 & D \\ 
AEGIS &      4711 &     2.1836 &      0.010 $\pm$      0.007 &       26.2 $\pm$        0.7 &       -3.5 $\pm$        0.8 & P \\ 
AEGIS &      6311 &     2.1878 &      0.200 $\pm$      0.008 &       -4.3 $\pm$        1.2 &       -9.2 $\pm$        1.1 & D \\ 
AEGIS &      6569 &     2.0857 &      0.080 $\pm$      0.008 &        7.9 $\pm$        5.2 &       -6.0 $\pm$        1.7 & C \\ 
AEGIS &     10471 &     2.3736 &      0.330 $\pm$      0.011 &      -15.7 $\pm$        4.7 &  $>$      -11.0 & L \\ 
AEGIS &     10494 &     2.2963 &      0.435 $\pm$      0.007 &      -19.4 $\pm$        3.0 &       -5.7 $\pm$        1.4 & C \\ 
AEGIS &     12918 &     2.4348 &      0.100 $\pm$      0.008 &       -8.1 $\pm$        6.4 &      -10.7 $\pm$        3.2 & P \\ 
AEGIS &     14957 &     2.3013 &      0.080 $\pm$      0.008 &      -10.2 $\pm$        2.4 &      -10.1 $\pm$        2.5 & C \\ 
AEGIS &     16496 &     2.1694 &      0.260 $\pm$      0.007 &       18.2 $\pm$        4.0 &      -14.2 $\pm$        5.0 & P \\ 
AEGIS &     18543 &     2.1387 &      0.030 $\pm$      0.008 &        1.5 $\pm$        0.8 &       -5.9 $\pm$        0.4 & D \\ 
AEGIS &     20924 &     2.2650 &      0.310 $\pm$      0.014 &       18.1 $\pm$       24.6 &      -9.1 $\pm$        0.4 & D \\ 
AEGIS &     21675 &     2.4631 &      0.190 $\pm$      0.014 &       -9.1 $\pm$        2.2 &  $>$       -5.8 & L \\ 
AEGIS &     22931 &     2.2952 &      0.050 $\pm$      0.007 &        9.9 $\pm$        0.4 &       -5.0 $\pm$        0.5 & D \\ 
AEGIS &     23409 &     2.2946 &      0.090 $\pm$      0.008 &       -3.0 $\pm$        0.9 &       -7.7 $\pm$        1.7 & P \\ 
AEGIS &     24481 &     2.1334 &      0.240 $\pm$      0.012 &      -10.3 $\pm$        1.2 &      -11.3 $\pm$        2.9 & P \\ 
AEGIS &     25522 &     2.2295 &      0.120 $\pm$      0.009 &      -25.8 $\pm$        2.3 &       -8.9 $\pm$        0.8 & D \\ 
AEGIS &     25817 &     2.2870 &      0.070 $\pm$      0.007 &      -23.0 $\pm$        3.6 &      -10.1 $\pm$        0.7 & D \\ 
AEGIS &     27627 &     1.6742 &      0.150 $\pm$      0.006 &      -15.5 $\pm$        3.4 &       -9.1 $\pm$        2.7 & P \\ 
AEGIS &     27825 &     2.2926 &      0.340 $\pm$      0.014 &      -10.2 $\pm$        2.4 &      -11.5 $\pm$        2.9 & P \\ 
AEGIS &     28421 &     2.2928 &      0.370 $\pm$      0.007 &      -17.6 $\pm$        4.9 &      -12.9 $\pm$        2.9 & P \\ 
AEGIS &     28659 &     2.3120 &      0.250 $\pm$      0.010 &      -19.6 $\pm$        2.5 &       -9.4 $\pm$        1.5 & D \\ 
AEGIS &     28710 &     2.1851 &      0.215 $\pm$      0.016 &       -7.9 $\pm$        4.7 &  $>$      -10.7 & L \\ 
AEGIS &     29650 &     2.2684 &      0.220 $\pm$      0.011 &       25.0 $\pm$        2.2 &       -8.1 $\pm$        0.8 & D \\ 
AEGIS &     30074 &     2.2113 &      0.310 $\pm$      0.010 &       -7.0 $\pm$        1.3 &       -9.2 $\pm$        1.9 & P \\ 
AEGIS &     30278 &     2.2003 &      0.230 $\pm$      0.014 &       13.2 $\pm$       18.6 &  $>$       -7.0 & L \\ 
AEGIS &     32354 &     2.1328 &      0.090 $\pm$      0.013 &        3.0 $\pm$        1.1 &       -7.7 $\pm$        0.7 & D \\ 
AEGIS &     32638 &     2.4080 &      0.120 $\pm$      0.013 &        0.7 $\pm$        0.8 &       -8.7 $\pm$        0.7 & D \\ 
AEGIS &     33768 &     2.3248 &      0.390 $\pm$      0.046 &        0.3 $\pm$        3.6 &       -9.9 $\pm$        2.7 & P \\ 
AEGIS &     33808 &     2.2254 &      0.140 $\pm$      0.007 &      -13.0 $\pm$        1.1 &       -9.0 $\pm$        1.8 & P \\ 
AEGIS &     33942 &     2.1608 &      0.090 $\pm$      0.009 &        1.7 $\pm$        1.5 &       -5.3 $\pm$        0.8 & D \\ 
AEGIS &     34661 &     2.1320 &      0.090 $\pm$      0.006 &      -23.7 $\pm$        3.1 &      -11.7 $\pm$        1.2 & D \\ 
AEGIS &     34813 &     2.2328 &      0.070 $\pm$      0.007 &      -16.9 $\pm$        2.4 &       -8.6 $\pm$        1.2 & D \\ 
AEGIS &     36257 &     2.1307 &      0.170 $\pm$      0.010 &      -10.0 $\pm$        1.8 &       -9.2 $\pm$        2.0 & P \\ 
AEGIS &     36451 &     2.1334 &      0.390 $\pm$      0.024 &       71.1 $\pm$        2.8 &  $>$      -10.9 & L \\ 
AEGIS &     40851 &     2.2672 &      0.010 $\pm$      0.009 &        6.0 $\pm$        1.2 &       -7.3 $\pm$        0.8 & D \\ 
COSMOS &       241 &     2.3131 &      0.030 $\pm$      0.007 &        1.4 $\pm$        0.9 &       -4.5 $\pm$        1.1 & C \\ 
COSMOS &       541 &     2.0810 &      0.060 $\pm$      0.006 &       24.6 $\pm$        2.4 &       -5.2 $\pm$        1.9 & P \\ 
COSMOS &       964 &     2.2903 &      0.080 $\pm$      0.006 &      -13.3 $\pm$        4.9 &      -12.9 $\pm$        3.6 & P \\ 
COSMOS &      2207 &     2.0976 &      0.060 $\pm$      0.007 &       12.7 $\pm$        2.7 &  $>$       -5.8 & L \\ 
COSMOS &      2672 &     2.3074 &      0.230 $\pm$      0.010 &       48.3 $\pm$       12.0 &  $>$      -11.1 & L \\ 
COSMOS &      2786 &     2.2980 &      0.120 $\pm$      0.008 &      -16.1 $\pm$        4.9 &  $>$       -6.9 & L \\ 
COSMOS &      3112 &     2.3080 &      0.220 $\pm$      0.009 &      -12.2 $\pm$        1.9 &       -9.3 $\pm$        2.5 & P \\ 
COSMOS &      3185 &     2.1732 &      0.090 $\pm$      0.007 &      -17.8 $\pm$        3.9 &       -8.2 $\pm$        2.6 & P \\ 
COSMOS &      3324 &     2.3072 &      0.280 $\pm$      0.009 &       -9.5 $\pm$        3.5 &      -14.7 $\pm$        3.5 & P \\ 
COSMOS &      3626 &     2.3247 &      0.040 $\pm$      0.004 &      -13.7 $\pm$        0.3 &       -5.6 $\pm$        0.4 & D \\ 
COSMOS &      3666 &     2.0859 &      0.310 $\pm$      0.007 &       -4.1 $\pm$        1.2 &       -4.2 $\pm$        1.2 & P \\ 
COSMOS &      3974 &     2.0979 &      0.230 $\pm$      0.005 &       78.0 $\pm$        7.3 &  $>$       -8.7 & L \\ 
COSMOS &      4078 &     2.4409 &      0.080 $\pm$      0.005 &      -20.1 $\pm$        6.2 &      -15.4 $\pm$        4.7 & P \\ 
COSMOS &      4156 &     2.1898 &      0.040 $\pm$      0.007 &       21.7 $\pm$        1.9 &       -7.2 $\pm$        1.9 & P \\ 
COSMOS &      4441 &     2.2243 &      0.050 $\pm$      0.006 &       -0.7 $\pm$        3.8 &       -5.1 $\pm$        1.6 & C \\ 
COSMOS &      4446 &     2.1970 &      0.080 $\pm$      0.005 &        8.7 $\pm$        0.9 &       -5.8 $\pm$        1.4 & P \\ 
COSMOS &      4497 &     2.4413 &      0.250 $\pm$      0.009 &      -18.2 $\pm$        5.3 &      -11.9 $\pm$        2.3 & P \\ 
COSMOS &      4930 &     2.2265 &      0.300 $\pm$      0.010 &       -9.0 $\pm$        4.5 &       -9.2 $\pm$        2.4 & P \\ 
COSMOS &      4945 &     2.0813 &      0.050 $\pm$      0.009 &        2.5 $\pm$        1.1 &  $>$       -3.9 & L \\ 
COSMOS &      4962 &     2.1725 &      0.050 $\pm$      0.007 &       -6.5 $\pm$        3.6 &  $>$       -7.3 & L \\ 
COSMOS &      5107 &     2.1443 &      0.200 $\pm$      0.009 &      -12.1 $\pm$        3.5 &       -6.1 $\pm$        1.7 & P \\ 
COSMOS &      5462 &     2.5221 &      0.270 $\pm$      0.007 &      -16.0 $\pm$        4.2 &       -8.0 $\pm$        2.6 & C \\ 
COSMOS &      5571 &     2.2779 &      0.240 $\pm$      0.010 &       -4.8 $\pm$        0.8 &       -5.3 $\pm$        0.8 & D \\ 
COSMOS &      5686 &     2.0956 &      0.260 $\pm$      0.011 &       -7.1 $\pm$        1.6 &       -8.1 $\pm$        2.9 & P \\ 
COSMOS &      5814 &     2.1266 &      0.290 $\pm$      0.008 &       -7.0 $\pm$        1.7 &       -7.6 $\pm$        1.0 & D \\ 
COSMOS &      5901 &     2.3962 &      0.200 $\pm$      0.012 &      -18.8 $\pm$        1.6 &       -9.6 $\pm$        2.0 & P \\ 
COSMOS &      6179 &     1.8506 &      0.420 $\pm$      0.006 &      -12.1 $\pm$        5.6 &      -15.8 $\pm$        5.2 & P \\ 
COSMOS &      6283 &     2.2238 &      0.070 $\pm$      0.005 &      -28.7 $\pm$        2.0 &       -8.6 $\pm$        0.9 & D \\ 
COSMOS &      6379 &     2.4382 &      0.090 $\pm$      0.006 &        9.8 $\pm$        2.2 &  $>$       -3.4 & L \\ 
COSMOS &      6417 &     2.0998 &      0.070 $\pm$      0.007 &      -22.0 $\pm$        2.6 &       -5.5 $\pm$        1.3 & P \\ 
COSMOS &      6817 &     2.0944 &      0.110 $\pm$      0.009 &      -11.1 $\pm$        3.6 &  $>$       -5.1 & L \\ 
COSMOS &      6826 &     2.4351 &      0.200 $\pm$      0.010 &        3.3 $\pm$        4.7 &  $>$       -7.7 & L \\ 
COSMOS &      6963 &     2.3014 &      0.340 $\pm$      0.004 &       22.3 $\pm$        2.5 &  $>$       -5.9 & L \\ 
COSMOS &      7430 &     1.9248 &      0.100 $\pm$      0.006 &      -15.0 $\pm$        9.2 &  $>$       -7.9 & L \\ 
COSMOS &      7735 &     2.4399 &      0.100 $\pm$      0.005 &       -3.2 $\pm$        1.2 &       -7.3 $\pm$        0.8 & D \\ 
COSMOS &      7883 &     2.1529 &      0.060 $\pm$      0.005 &       18.4 $\pm$        1.2 &       -7.6 $\pm$        2.8 & P \\ 
COSMOS &      8081 &     2.1633 &      0.070 $\pm$      0.008 &      -39.1 $\pm$       10.9 &  $>$       -6.0 & L \\ 
COSMOS &      8515 &     2.4537 &      0.060 $\pm$      0.003 &        6.0 $\pm$        1.2 &       -4.2 $\pm$        0.6 & D \\ 
COSMOS &      8540 &     2.0912 &      0.060 $\pm$      0.007 &      -15.9 $\pm$        2.1 &       -6.1 $\pm$        1.6 & C \\ 
COSMOS &      9044 &     2.1988 &      0.150 $\pm$      0.009 &      -15.4 $\pm$        1.8 &       -9.3 $\pm$        1.1 & D \\ 
COSMOS &      9251 &     2.2426 &      0.050 $\pm$      0.009 &       28.6 $\pm$        4.8 &       -8.6 $\pm$        3.0 & C \\ 
COSMOS &     10066 &     2.4133 &      0.090 $\pm$      0.005 &        5.7 $\pm$        2.1 &      -12.8 $\pm$        4.1 & P \\ 
COSMOS &     10143 &     2.3776 &      0.060 $\pm$      0.004 &       14.7 $\pm$        0.7 &       -4.5 $\pm$        1.0 & P \\ 
COSMOS &     10235 &     2.0999 &      0.200 $\pm$      0.008 &      -17.6 $\pm$        8.1 &       -9.2 $\pm$        2.4 & P \\ 
COSMOS &     10280 &     2.1945 &      0.060 $\pm$      0.005 &       -8.2 $\pm$        1.5 &       -5.1 $\pm$        0.6 & D \\ 
COSMOS &     10835 &     2.4139 &      0.100 $\pm$      0.005 &       25.7 $\pm$       19.8 &  $>$      -13.9 & L \\ 
COSMOS &     11443 &     2.4561 &      0.080 $\pm$      0.006 &        0.8 $\pm$        2.1 &      -11.3 $\pm$        3.6 & P \\ 
COSMOS &     11530 &     2.0969 &      0.030 $\pm$      0.009 &       57.2 $\pm$        1.4 &       -2.5 $\pm$        0.3 & D \\ 
COSMOS &     12577 &     2.5365 &      0.350 $\pm$      0.004 &      -22.2 $\pm$        1.7 &      -10.8 $\pm$        3.8 & P \\ 
COSMOS &     13101 &     1.9727 &      0.200 $\pm$      0.006 &       -6.2 $\pm$        3.5 &  $>$      -11.6 & L \\ 
COSMOS &     13299 &     2.3089 &      0.310 $\pm$      0.009 &      -39.1 $\pm$       17.3 &  $>$       -9.8 & L \\ 
COSMOS &     13364 &     2.1508 &      0.060 $\pm$      0.006 &       10.9 $\pm$        1.5 &      -11.2 $\pm$        3.7 & P \\ 
COSMOS &     16545 &     2.2750 &      0.070 $\pm$      0.005 &       -4.3 $\pm$        0.9 &       -6.9 $\pm$        0.4 & D \\ 
COSMOS &     19439 &     2.4663 &      0.210 $\pm$      0.013 &       13.0 $\pm$        2.1 &       -9.1 $\pm$        1.2 & D \\ 
COSMOS &     19712 &     2.4863 &      0.350 $\pm$      0.014 &      -17.1 $\pm$        5.8 &  $>$      -28.5 & L \\ 
COSMOS &     19985 &     2.1882 &      0.270 $\pm$      0.011 &      -47.3 $\pm$        1.7 &      -12.0 $\pm$        0.5 & D \\ 
COSMOS &     20062 &     2.1857 &      0.260 $\pm$      0.011 &       -3.9 $\pm$        0.5 &      -11.2 $\pm$        0.6 & D \\ 
COSMOS &     21780 &     2.4718 &      0.060 $\pm$      0.005 &      -10.3 $\pm$        0.8 &       -5.8 $\pm$        0.9 & D \\ 
COSMOS &     21955 &     2.4676 &      0.230 $\pm$      0.012 &      -22.2 $\pm$        5.2 &      -13.8 $\pm$        4.0 & P \\ 
COSMOS &     24020 &     2.0923 &      0.080 $\pm$      0.006 &       10.7 $\pm$        2.3 &       -6.9 $\pm$        2.4 & P \\ 
COSMOS &     25322 &     2.5188 &      0.050 $\pm$      0.005 &        7.8 $\pm$        0.8 &       -6.7 $\pm$        1.4 & P \\ 
COSMOS &     26073 &     2.2235 &      0.350 $\pm$      0.015 &       21.2 $\pm$       12.3 &  $>$      -46.5 & L \\ 
COSMOS &     27120 &     2.4784 &      0.160 $\pm$      0.006 &       -1.9 $\pm$        0.6 &       -6.6 $\pm$        0.7 & D \\ 
COSMOS &     27216 &     2.4225 &      0.030 $\pm$      0.005 &       25.9 $\pm$        3.4 &      -10.0 $\pm$        2.8 & P \\ 
COSMOS &     27906 &     2.1961 &      0.050 $\pm$      0.005 &       -7.5 $\pm$        2.0 &       -6.0 $\pm$        1.2 & P \\ 
COSMOS &     28258 &     2.4732 &      0.205 $\pm$      0.010 &      -20.1 $\pm$        5.2 &       -9.7 $\pm$        3.0 & P \\ 
GOODS-N &     10596 &     2.2135 &      0.150 $\pm$      0.009 &       -4.7 $\pm$        0.6 &       -7.6 $\pm$        0.7 & D \\ 
GOODS-N &     10645 &     2.1796 &      0.190 $\pm$      0.010 &      -13.0 $\pm$        5.3 &      -13.6 $\pm$        1.5 & D \\ 
GOODS-N &     12157 &     2.2765 &      0.195 $\pm$      0.020 &       -2.8 $\pm$        2.3 &      -10.6 $\pm$        1.3 & D \\ 
GOODS-N &     12345 &     2.2721 &      0.220 $\pm$      0.010 &       -9.8 $\pm$        2.6 &       -9.9 $\pm$        0.5 & D \\ 
GOODS-N &     12980 &     2.2697 &      0.090 $\pm$      0.006 &      -20.4 $\pm$        5.2 &       -8.2 $\pm$        2.3 & P \\ 
GOODS-N &     15186 &     2.4139 &      0.090 $\pm$      0.010 &      -44.2 $\pm$       30.0 &  $>$      -10.7 & L \\ 
GOODS-N &     16351 &     1.6511 &      0.060 $\pm$      0.007 &       24.0 $\pm$        3.8 &  $>$       -5.7 & L \\ 
GOODS-N &     17530 &     2.2064 &      0.100 $\pm$      0.006 &       15.0 $\pm$        1.3 &  $>$       -1.8 & L \\ 
GOODS-N &     17714 &     2.2349 &      0.070 $\pm$      0.006 &      -17.7 $\pm$        2.2 &       -7.2 $\pm$        1.4 & P \\ 
GOODS-N &     19067 &     2.2829 &      0.070 $\pm$      0.006 &       -3.8 $\pm$        1.2 &       -8.2 $\pm$        2.4 & P \\ 
GOODS-N &     19350 &     2.2367 &      0.070 $\pm$      0.008 &      -18.8 $\pm$        1.7 &       -6.7 $\pm$        1.0 & D \\ 
GOODS-N &     19654 &     2.5519 &      0.060 $\pm$      0.013 &       -6.0 $\pm$        1.3 &      -15.1 $\pm$        5.3 & P \\ 
GOODS-N &     20924 &     2.5511 &      0.150 $\pm$      0.008 &      -20.1 $\pm$        1.9 &  $>$      -15.7 & L \\ 
GOODS-N &     21279 &     2.4197 &      0.360 $\pm$      0.010 &      -38.6 $\pm$        5.8 &  $>$       -8.6 & L \\ 
GOODS-N &     21617 &     2.2062 &      0.200 $\pm$      0.009 &        1.7 $\pm$        0.6 &       -5.7 $\pm$        0.5 & D \\ 
GOODS-N &     21845 &     2.5509 &      0.190 $\pm$      0.009 &       -2.1 $\pm$        1.1 &       -6.1 $\pm$        1.6 & P \\ 
GOODS-N &     22235 &     2.4298 &      0.070 $\pm$      0.005 &       -1.6 $\pm$        1.2 &       -6.4 $\pm$        0.5 & D \\ 
GOODS-N &     22487 &     2.4205 &      0.310 $\pm$      0.010 &      -13.8 $\pm$        5.2 &       -6.9 $\pm$        2.1 & C \\ 
GOODS-N &     22669 &     2.1340 &      0.140 $\pm$      0.011 &      -13.4 $\pm$        0.9 &       -6.5 $\pm$        0.6 & D \\ 
GOODS-N &     23344 &     2.4839 &      0.370 $\pm$      0.026 &      -23.5 $\pm$        6.9 &       -8.5 $\pm$        3.1 & P \\ 
GOODS-N &     23869 &     2.2438 &      0.290 $\pm$      0.011 &      -28.6 $\pm$        6.6 &  $>$       -6.7 & L \\ 
GOODS-N &     24328 &     2.4072 &      0.070 $\pm$      0.005 &      -32.9 $\pm$        0.8 &       -7.5 $\pm$        0.5 & D \\ 
GOODS-N &     24825 &     2.3347 &      0.060 $\pm$      0.005 &       -4.1 $\pm$        1.7 &  $>$       -2.8 & L \\ 
GOODS-N &     24846 &     2.1872 &      0.040 $\pm$      0.005 &        7.8 $\pm$        1.7 &       -6.4 $\pm$        1.7 & P \\ 
GOODS-N &     25142 &     2.4691 &      0.240 $\pm$      0.011 &      -40.6 $\pm$        6.0 &       -9.2 $\pm$        2.6 & C \\ 
GOODS-N &     25688 &     2.3748 &      0.070 $\pm$      0.005 &        0.6 $\pm$        0.4 &       -7.5 $\pm$        0.5 & D \\ 
GOODS-N &     26621 &     2.3055 &      0.090 $\pm$      0.005 &      -31.0 $\pm$        3.8 &       -7.0 $\pm$        1.5 & P \\ 
GOODS-N &     27035 &     2.4218 &      0.080 $\pm$      0.005 &      -22.1 $\pm$        4.5 &       -7.6 $\pm$        2.1 & P \\ 
GOODS-N &     28237 &     2.2266 &      0.080 $\pm$      0.008 &       -3.1 $\pm$        0.8 &  $>$       -2.3 & L \\ 
GOODS-N &     28599 &     1.6871 &      0.090 $\pm$      0.005 &        3.6 $\pm$        4.6 &       -6.4 $\pm$        2.1 & C \\ 
GOODS-N &     28846 &     2.4720 &      0.090 $\pm$      0.006 &        3.1 $\pm$        1.8 &       -5.5 $\pm$        1.9 & P \\ 
GOODS-N &     29743 &     2.1867 &      0.210 $\pm$      0.009 &       13.0 $\pm$        2.2 &       -8.4 $\pm$        1.0 & D \\ 
GOODS-N &     30053 &     2.2452 &      0.320 $\pm$      0.011 &        6.5 $\pm$        5.0 &      -12.8 $\pm$        3.6 & P \\ 
GOODS-N &     32526 &     2.4088 &      0.100 $\pm$      0.010 &      -23.7 $\pm$       15.3 &  $>$      -10.8 & L \\ 
GOODS-S &     31344 &     2.3237 &      0.230 $\pm$      0.007 &       -6.0 $\pm$        1.2 &       -9.8 $\pm$        1.0 & D \\ 
GOODS-S &     31854 &     2.4250 &      0.080 $\pm$      0.005 &       -5.0 $\pm$        1.2 &       -4.2 $\pm$        1.3 & C \\ 
GOODS-S &     32837 &     2.0608 &      0.065 $\pm$      0.006 &       -4.0 $\pm$        0.9 &       -5.7 $\pm$        1.2 & P \\ 
GOODS-S &     33248 &     2.3245 &      0.090 $\pm$      0.008 &      -23.8 $\pm$        5.8 &  $>$       -4.3 & L \\ 
GOODS-S &     35178 &     2.4084 &      0.100 $\pm$      0.005 &      -11.3 $\pm$        3.3 &       -7.8 $\pm$        2.6 & P \\ 
GOODS-S &     35705 &     2.3234 &      0.150 $\pm$      0.008 &       -0.5 $\pm$        1.2 &       -6.3 $\pm$        1.6 & P \\ 
GOODS-S &     35779 &     2.2536 &      0.050 $\pm$      0.005 &        5.2 $\pm$        1.4 &       -7.1 $\pm$        2.4 & P \\ 
GOODS-S &     36705 &     2.3064 &      0.130 $\pm$      0.006 &        3.5 $\pm$        0.4 &       -7.6 $\pm$        0.4 & D \\ 
GOODS-S &     37988 &     2.2008 &      0.330 $\pm$      0.009 &      -14.6 $\pm$        1.6 &  $>$       -6.6 & L \\ 
GOODS-S &     38116 &     2.1968 &      0.260 $\pm$      0.009 &      -21.7 $\pm$        4.9 &       -7.8 $\pm$        1.9 & P \\ 
GOODS-S &     38559 &     2.1939 &      0.080 $\pm$      0.005 &        2.8 $\pm$        0.9 &       -5.5 $\pm$        0.9 & D \\ 
GOODS-S &     39198 &     2.5789 &      0.030 $\pm$      0.005 &       12.3 $\pm$        0.5 &       -5.3 $\pm$        1.3 & P \\ 
GOODS-S &     39713 &     2.1546 &      0.050 $\pm$      0.006 &       10.2 $\pm$        1.1 &       -7.4 $\pm$        0.9 & D \\ 
GOODS-S &     40218 &     2.4508 &      0.080 $\pm$      0.005 &       -8.0 $\pm$        0.6 &       -7.9 $\pm$        0.4 & D \\ 
GOODS-S &     40679 &     2.4087 &      0.280 $\pm$      0.009 &      -10.9 $\pm$        3.6 &       -9.1 $\pm$        2.8 & P \\ 
GOODS-S &     40768 &     2.3035 &      0.190 $\pm$      0.010 &      -36.0 $\pm$        1.5 &      -14.4 $\pm$        0.7 & D \\ 
GOODS-S &     41547 &     2.5451 &      0.060 $\pm$      0.005 &      -12.4 $\pm$        0.5 &       -7.5 $\pm$        0.3 & D \\ 
GOODS-S &     42363 &     2.1411 &      0.210 $\pm$      0.009 &       -6.2 $\pm$        1.1 &      -14.2 $\pm$        0.8 & D \\ 
GOODS-S &     42809 &     2.2494 &      0.020 $\pm$      0.005 &       12.9 $\pm$        1.0 &       -3.6 $\pm$        0.9 & P \\ 
GOODS-S &     45180 &     2.2858 &      0.090 $\pm$      0.005 &        7.7 $\pm$        2.2 &       -8.8 $\pm$        3.1 & P \\ 
GOODS-S &     45531 &     2.3116 &      0.040 $\pm$      0.005 &       13.2 $\pm$        0.5 &       -5.4 $\pm$        0.5 & D \\ 
GOODS-S &     46938 &     2.3325 &      0.060 $\pm$      0.003 &       22.6 $\pm$        0.9 &       -6.5 $\pm$        0.2 & D \\ 

   \enddata
\tablecomments{The EW values listed are in the rest-frame. The LIS EW represents the sum of \textrm{Si}\textsc{ii}$\lambda1260$,  \textrm{O}\textsc{i}$\lambda1302$+\textrm{Si}\textsc{ii}$\lambda1304$,  and \textrm{C}\textsc{ii}$\lambda1334$ \textrm{C}\textsc{iii}. For LIS non-detections, a $3\sigma$ limit is reported. The LIS category denotes how the LIS EW was calculated: `D' represents the cases where all 3 lines are individually detected; `P' represents where only 1 or 2 LIS lines are detected at the $\geqslant3\sigma$ level, and the LIS EW was inferred based on the method described in Section \ref{sec:lis}; `C' represents where none of the LIS lines were individually detected, but the combined line $S/N$ is $\geqslant3$; `L' represents where none of the LIS lines were individually detected and the combined line $S/N$ is $<3$, therefore a limit is reported.}
\label{tab:galprop}
\end{deluxetable*}

\section{Results}
\label{sec:results}

\begin{figure}
\includegraphics[width=1.0\linewidth]{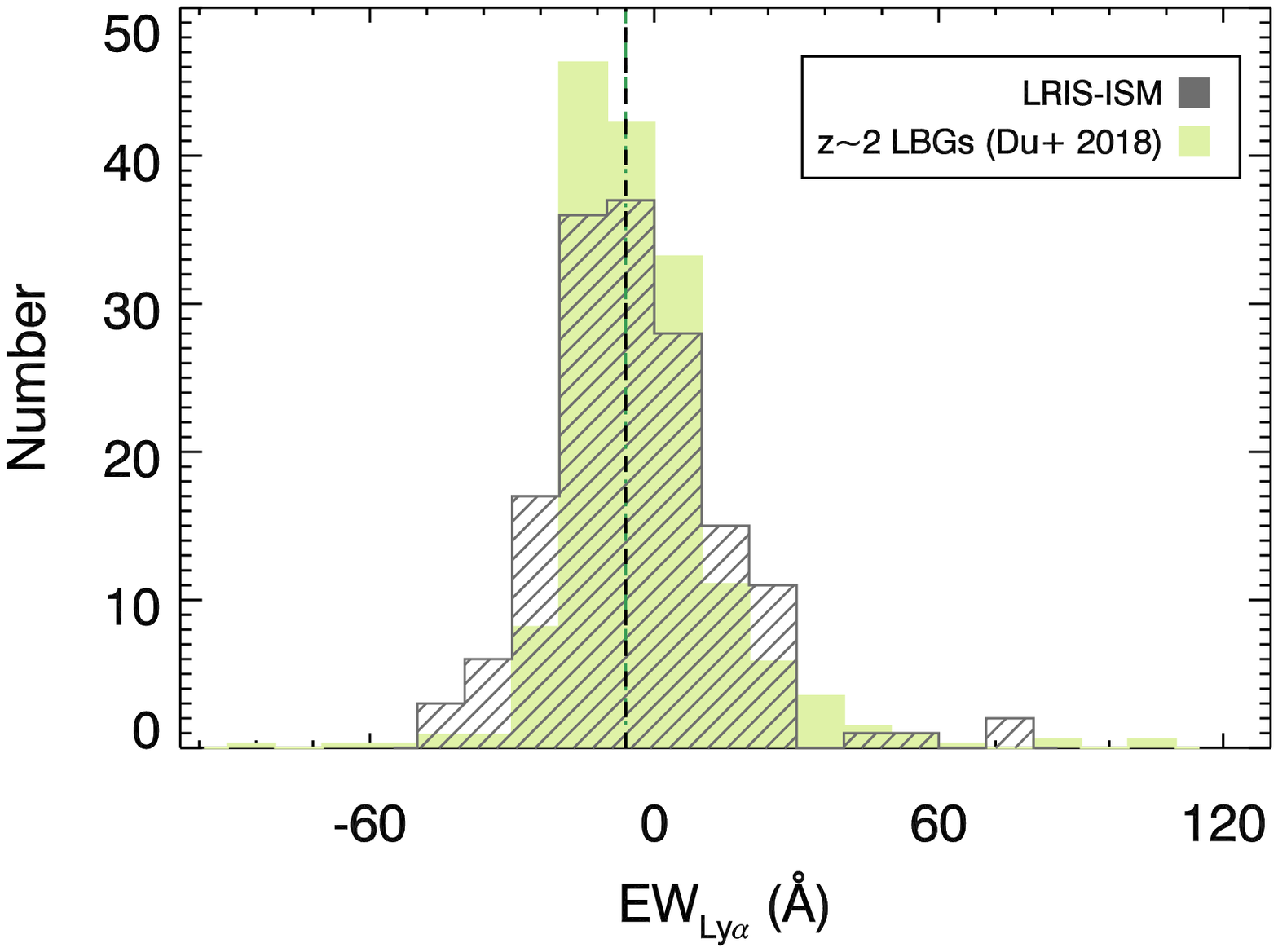}
\includegraphics[width=1.0\linewidth]{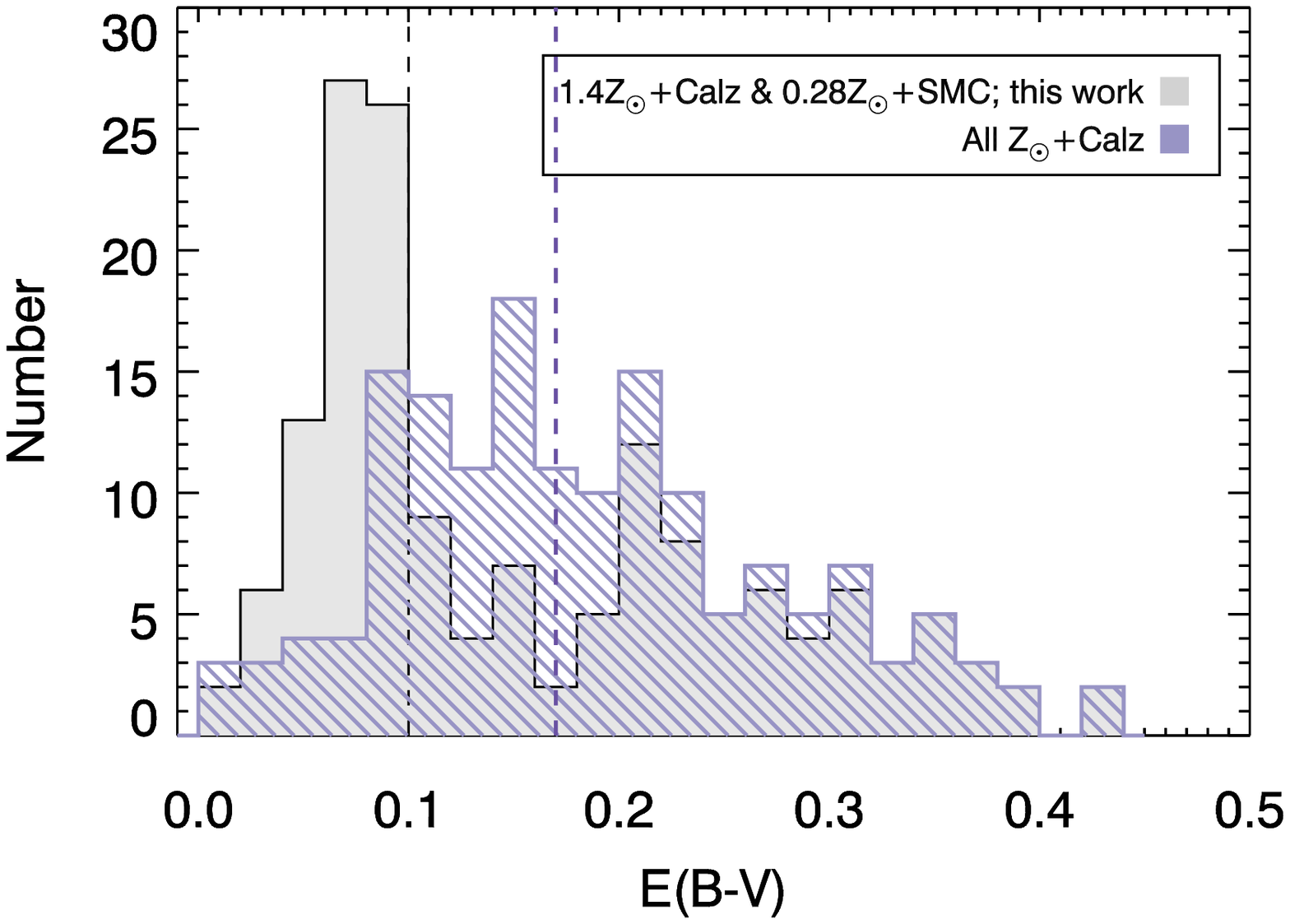}
\caption{\textbf{Top:} Rest-frame Ly$\alpha$ EW distribution. The filled green bar indicates the $z\sim2$ LBG sample (538 objects) presented in \citet{Du2018}, normalized to the same total number as the LRIS-ISM sample (157 objects; shaded gray bar). The median EWs are -6.0 $\mbox{\AA}$ and -6.1 $\mbox{\AA}$ for the LRIS-ISM and the $z\sim2$ LBG samples, respectively, shown with a dashed black line for the former and a dash-dotted green line for the latter. \textbf{Bottom:} $E(B-V)$ distribution. The filled gray bar shows the distribution of $E(B-V)$ estimated based on the 1.4 $Z_{\sun}$+Calzetti (0.28 $Z_{\sun}$+SMC) model for objects above (below) stellar mass log$(M_{*}/M_{\sun})=10.04$, following the stellar population modeling procedures described in Section \ref{sec:sed}. The median $E(B-V)$ is 0.10, as denoted by the vertical dashed black line. In comparison, the $E(B-V)$ distribution derived by fitting all objects with the 1.4 $Z_{\sun}$+Calzetti model is plotted in shaded purple bar. The corresponding median $E(B-V)$ is 0.17 (purple dashed line).} 
\label{fig:ew_hist}
\end{figure}

\subsection{Ly$\alpha$ and $E(B-V)$ Distributions}
\label{sec:ew}

Rest-frame EW$_{Ly\alpha}$ and $E(B-V)$ were measured using the methods described in Section \ref{sec:measure} for all objects in the LRIS-ISM sample. The individual Ly$\alpha$ measurements are listed in Table \ref{tab:galprop}. As shown in the upper panel of Figure \ref{fig:ew_hist}, the Ly$\alpha$ EW ranges from $-47.3 \mbox{\AA}$ to 78.0 $\mbox{\AA}$ with a median of $-6.0 \mbox{\AA}$. The median EW$_{Ly\alpha}$ of the $H$-band-selected LRIS-ISM sample is similar to that of the $z\sim2$ UV-selected Lyman Break Galaxy (LBG) sample presented in \citet{Du2018}, and the EW$_{Ly\alpha}$ distribution of the LRIS-ISM sample is slightly wider (the standard deviations are $19.1 \mbox{\AA}$ and $18.0 \mbox{\AA}$, respectively, for the LRIS-ISM and the $z\sim2$ LBG EW$_{Ly\alpha}$ distributions).

As for dust attenuation, the derived $E(B-V)$ value for the LRIS-ISM sample ranges from 0.01 to 0.435, with a median of 0.10. This median $E(B-V)$ value is very close to that of the $z\sim2$ LBG sample (median $E(B-V)=0.09$) in \citet{Du2018}, in which similar SED modeling approaches were adopted. We note that the 1.4 $Z_{\sun}$+Calzetti model outputs a systematically higher $E(B-V)$ than the 0.28 $Z_{\sun}$+SMC for the same galaxy SED, hence the peak shown near $E(B-V)\sim0.08-0.10$ in Figure \ref{fig:ew_hist} is caused by the adoption of the 0.28 $Z_{\sun}$+SMC model (smaller $E(B-V)$) for lower-mass galaxies. Assuming the 1.4 $Z_{\sun}$+Calzetti model for all objects in the sample would result in a flatter distribution, with a median $E(B-V)=0.17$. However, as justified in Section \ref{sec:sed}, we believe that the 0.28 $Z_{\sun}$+SMC model better characterizes the lower-mass (log$(M_{*}/M_{\sun})<10.04$) galaxies in our sample. Additionally, as the $E(B-V)$ values derived from the 1.4 $Z_{\sun}$+Calzetti and 0.28 $Z_{\sun}$+SMC models are tightly correlated, adopting a combination of these two models would not considerably change the dust extinction among the galaxies in a relative sense (i.e., galaxies with a higher than average $E(B-V)$ output by the 1.4 $Z_{\sun}$+Calzetti model still have a relatively high $E(B-V)$ derived from the 0.28 $Z_{\sun}$+SMC model). As a result, the qualitative trends between $E(B-V)$ and both EW$_{Ly\alpha}$ and EW$_{LIS}$ (Section \ref{sec:trends}) should not be significantly affected. 

\subsection{Relations Among EW$_{Ly\alpha}$, EW$_{LIS}$, And $E(B-V)$}
\label{sec:trends}

\begin{figure}
\includegraphics[width=1.0\linewidth]{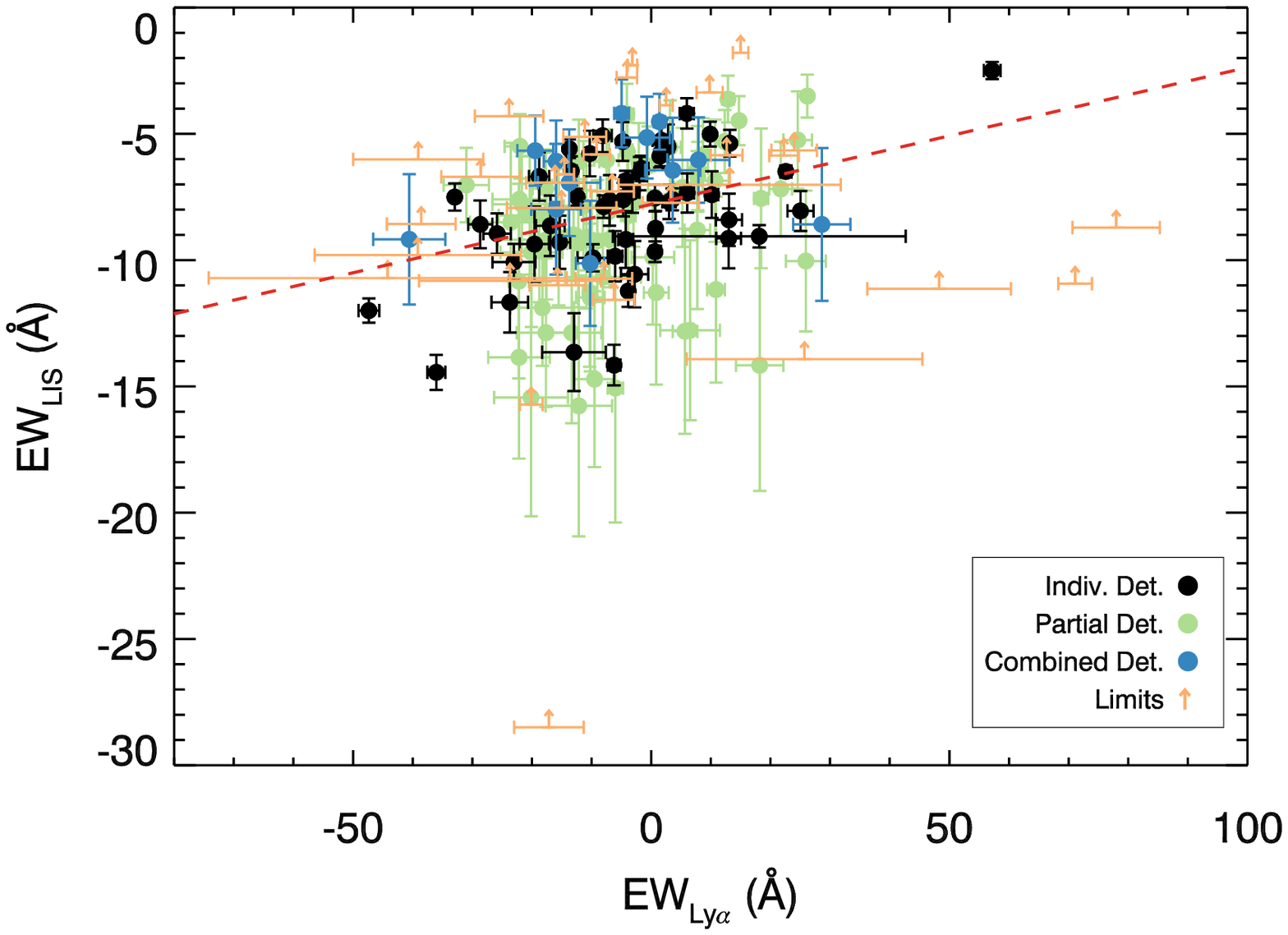}
\includegraphics[width=1.0\linewidth]{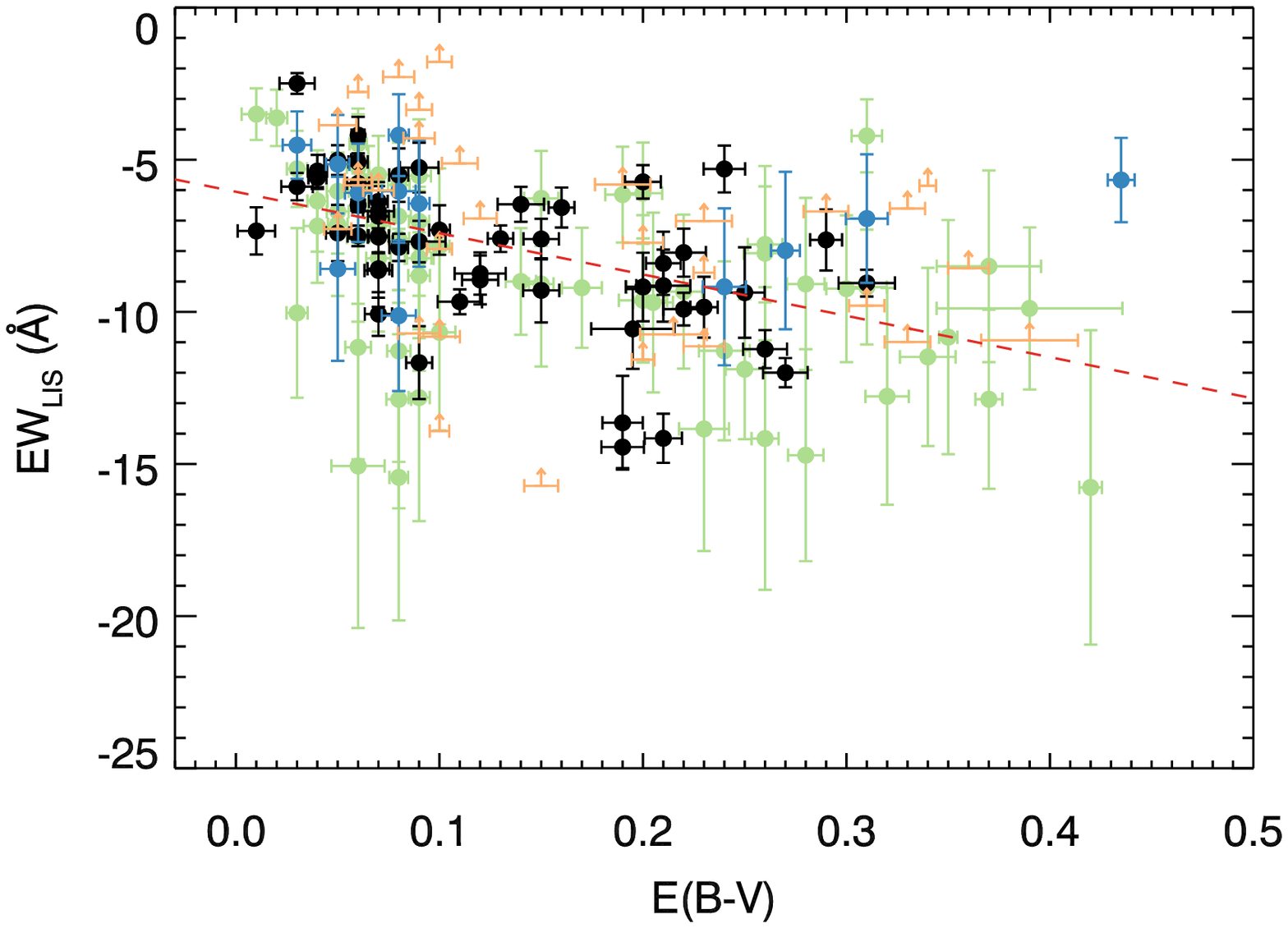}
\includegraphics[width=1.0\linewidth]{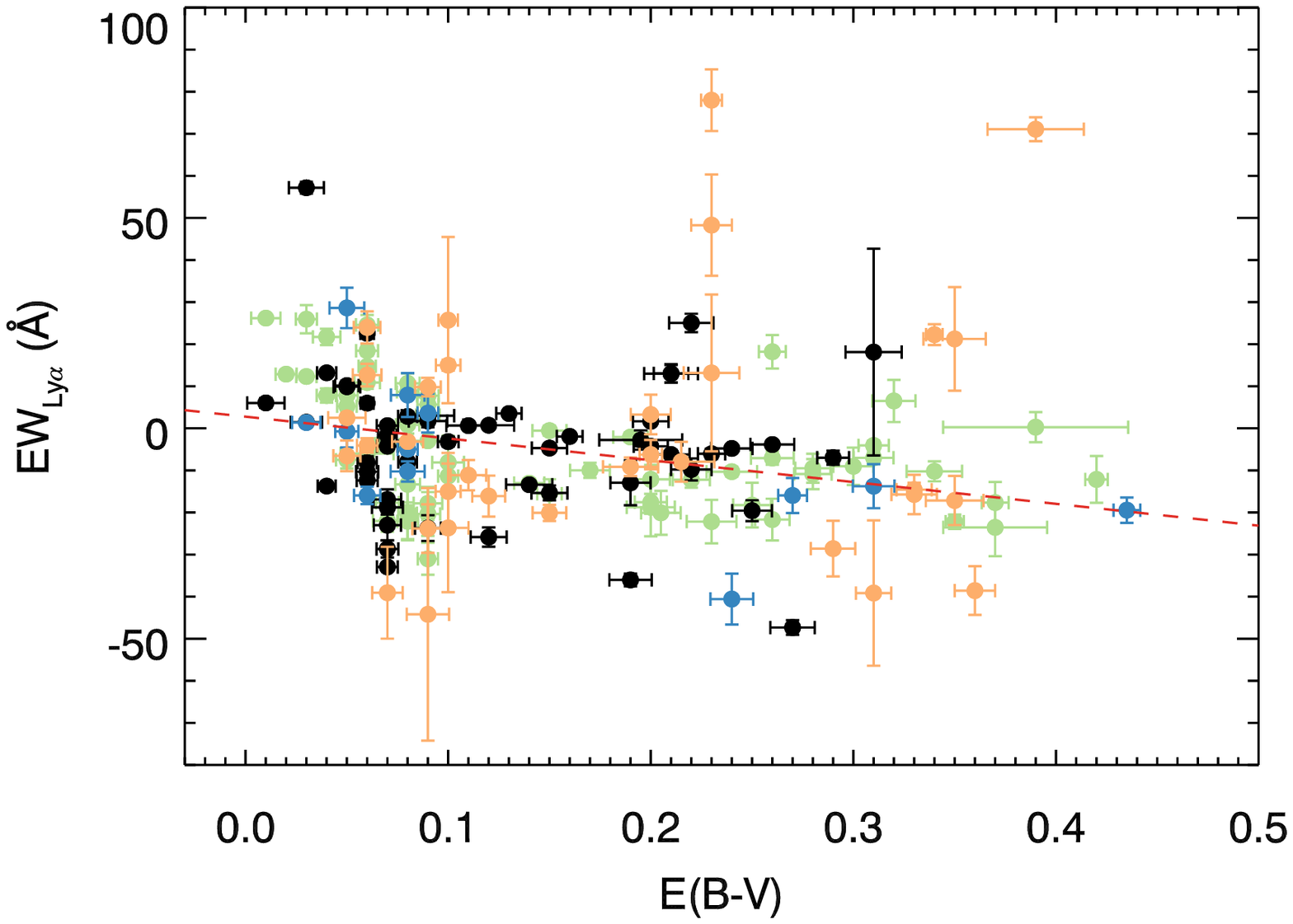}
\caption{Correlations among EW$_{Ly\alpha}$, EW$_{LIS}$, and $E(B-V)$ using individual measurements from 157 galaxies in the LRIS-ISM sample. \textbf{Top:} EW$_{LIS}$ vs. EW$_{Ly\alpha}$. The black, green, and blue circles show the objects with individual detections (51 objects), partial detections (61 objects), and combined detections (12 objects) as classified in Section \ref{sec:lis}. The 1$\sigma$ error bar is plotted for detections. The orange upward-pointing arrows denote the 3$\sigma$ limit on EW$_{LIS}$ for 33 objects where the LIS lines are not significantly detected. The limits are plotted as lower limits as we define absorption line EW as negative. The dashed red line marks the best-fit linear regression, returned by LINMIX$\_$ERR (see Section \ref{sec:trends}. \textbf{Middle:} EW$_{LIS}$ vs. $E(B-V)$. Color coding of the symbols is the same as in the top left panel. \textbf{Bottom:} EW$_{Ly\alpha}$ vs. $E(B-V)$. Color coding of the symbols is the same as in the top panels, except that the EW$_{LIS}$ limits are plotted as circles, as those 33 objects all have measured EW$_{Ly\alpha}$ and $E(B-V)$.} 
\label{fig:corr}
\end{figure}

One key aspect of this study is to quantify the relative tightness of the relations among EW$_{Ly\alpha}$, EW$_{LIS}$, and $E(B-V)$ using measurements from {\it individual} galaxies. To avoid potential bias introduced by selection effects, we conducted statistical analysis using the entire LRIS-ISM sample (157 objects), which includes 33 limits in EW$_{LIS}$. With the measurements of EW$_{Ly\alpha}$, EW$_{LIS}$, and $E(B-V)$ obtained in Section \ref{sec:measure}, we plot the relations among these 3 parameters in Figure \ref{fig:corr}. To parameterize the correlations, we used an IDL package LINMIX$\_$ERR \citep{Kelly2007} for performing a linear regression between each pair of observables. The linear function takes the form y $=$ intercept $+$ slope $*$ x + $\epsilon$, where $\epsilon$ represents the intrinsic random scatter in the regression and is a normal distribution with a zero mean. LINMIX$\_$ERR is an ideal program for dealing with complex data like ours, as it not only takes into account the measurement uncertainties in both variables but also allows censored data (i.e., non-detections). The program adopts a Bayesian approach for calculating the linear regression, and each parameter (such as intercept, slope, and intrinsic scatter) is returned as an array of 200 draws from its associated posterior distribution. We report the regression coefficients in Table \ref{tab:reg}, where the reported values and error bars were determined based on the median and the standard deviation of the 200 draws of the respective parameter.

The regression coefficients in Table \ref{tab:reg} suggest that EW$_{Ly\alpha}$ is larger in galaxies with smaller EW$_{LIS}$ and smaller $E(B-V)$, and EW$_{LIS}$ increases with increasing $E(B-V)$. These qualitative findings are consistent with previous work at similar redshifts using galaxy composite spectra \citep[e.g.,][]{Shapley2003,Du2018}. 
However, our results represent the first such analysis of the mutual correlations among EW$_{Ly\alpha}$, EW$_{LIS}$, and $E(B-V)$ based on {\it individual} measurements at high redshift, enabling an investigation of the scatter in these key relationships.

While the ``intrinsic scatter" term can be useful in quantifying the tightness of the correlations, the values returned by LINMIX$\_$ERR describe the deviation in the y-axis and can only be used for direct comparison if the dependent variable is the same. Given that EW$_{Ly\alpha}$ and EW$_{LIS}$ can be the dependent variable in different correlations, and that their dynamic ranges and distributions are drastically different, we therefore seek alternative programs for performing statistical analysis on the correlations (see Section \ref{sec:corr}).

\begin{table*}
\centering
\begin{threeparttable}
  \caption{Linear Regression Coefficients}
  \label{tab:reg}
{\renewcommand{\arraystretch}{1.5}
  \begin{tabular}{cccc}
  \hline
  \hline
    Correlation & Intercept & Slope & Intrinsic Scatter \tnote{1} \\
    & (\mbox{\AA}) & & (\mbox{\AA}) \\
  \hline
  \multicolumn{4}{c}{Entire Sample (157 objects)} \\
  \hline
  EW$_{LIS}$ vs. $E(B-V)$ & $-$6.086 $\pm$ 0.397 & $-$12.871 $\pm$ 2.511 & 1.816 $\pm$     0.191 \\ 
 \hline
 EW$_{LIS}$ vs. EW$_{Ly\alpha}$ & $-$7.759 $\pm$ 0.198 & 0.062 $\pm$ 0.012 & 1.916 $\pm$     0.170 \\ 
 \hline
 EW$_{Ly\alpha}$ vs. $E(B-V)$ &  3.037 $\pm$ 2.353 & $-$53.813 $\pm$ 12.682 & 14.649 $\pm$ 0.913 \\ 
 \hline
 \end{tabular}}
\begin{tablenotes}
\item[1] The linear regression assumes a form: y $=$ intercept $+$ slope $*$ x + $\epsilon$, where $\epsilon$ follows a normal distribution with zero mean and  variance equal to the square of the intrinsic scatter. 
 \end{tablenotes}
\end{threeparttable}
\end{table*}

\subsection{Relative Strengths of the Key ISM/CGM Relations}
\label{sec:corr}

To utilize the entire sample and directly compare the relative tightness of the mutual correlations among EW$_{Ly\alpha}$, EW$_{LIS}$, and $E(B-V)$, we adopted the FORTRAN routine ASURV \citep{Feigelson1985,Isobe1986,Isobe1990,Lavalley1992} for performing survival analysis of censored data, which was written specifically to treat non-detections due to sensitivity limits. ASURV offers various statistical tests, out of which we chose to use the generalized Kendall's $\tau$ and Spearman's $\rho$ correlation coefficients for characterizing the tightness of the correlations. Both correlation coefficients assess statistical associations based on the ranks of the data, and yield almost identical results for our sample. The resulting Kendall's $\tau$ and Spearman's $\rho$ are listed in Table \ref{tab:corr}, along with the corresponding probability of a null hypothesis (i.e., that the data are uncorrelated).

One complication in using ASURV is that the routine does not consider measurement uncertainties on the variables, but all our measurements have associated 1$\sigma$ error-bars except for EW$_{LIS}$ non-detections (in which case a 3$\sigma$ limit was fed to ASURV). To account for the measurement uncertainties, for each correlation (EW$_{LIS}$ vs. EW$_{Ly\alpha}$, EW$_{Ly\alpha}$ vs. $E(B-V)$, or EW$_{LIS}$ vs. $E(B-V)$), we perturbed the detections in both variables by their individual error bars for 100 times, while keeping the 3$\sigma$ limit of the non-detections unchanged. The standard deviation of those 100 realizations on Kendall's $\tau$, Spearman's $\rho$, and the $P_{K}$ and $P_{SR}$ values was taken as the error bar on the respective parameters in Table \ref{tab:corr}, to reflect the uncertainty on the statistical analysis results introduced by measurement errors. 

The statistical tests show that EW$_{Ly\alpha}$, EW$_{LIS}$, and $E(B-V)$ are inter-correlated. All three correlations have a $>3\sigma$ significance, but the EW$_{LIS}$ vs. $E(B-V)$ relation is the strongest. The strength between EW$_{Ly\alpha}$ vs. EW$_{LIS}$ and EW$_{Ly\alpha}$ vs. $E(B-V)$ are comparable considering the uncertainty on the correlation coefficients, with the former being slightly weaker. We note that swapping the x- and y-axis observables (e.g., from EW$_{LIS}$ vs. $E(B-V)$ to $E(B-V)$ vs. EW$_{LIS}$) yields similar correlation coefficients and does not change the relative strength of the mutual correlations we examine here. Provided that the EW$_{Ly\alpha}$ vs. $E(B-V)$ relation does not contain any measurement limits but the other two relations do, we also tested whether excluding the limits would change the results. We performed similar statistical analyses using ASURV on the 124 objects with detected LIS features (including the individual, partial, and combined detections as described in Section \ref{sec:lis}). When considering detections only, the correlation between EW$_{LIS}$ and $E(B-V)$ is again the strongest, while that between EW$_{Ly\alpha}$ vs. EW$_{LIS}$ is the weakest. 
Therefore, including limits does not change the relative tightness of the three correlations. 

Our results suggest the most direct connection exists between the \textrm{H}\textsc{i} covering fraction and dust attenuation, as probed by EW$_{LIS}$ and $E(B-V)$, respectively. The EW$_{Ly\alpha}$ vs. $E(B-V)$ correlation is the second strongest, highlighting the preferential dust extinction of Ly$\alpha$ photons relative to continuum photons within the region covered by the LRIS spectroscopic slit. Alternatively, the observed EW$_{Ly\alpha}$ vs. $E(B-V)$ trend can be explained by more scattering of Ly$\alpha$ photons out of the spectroscopic slit by the higher \textrm{H}\textsc{i} covering fraction typically associated with higher $E(B-V)$. Finally, although the weakest among the three, the correlation between EW$_{LIS}$ vs. EW$_{Ly\alpha}$ point to the impact of resonant scattering on the emergent Ly$\alpha$ emission when Ly$\alpha$ photons travel through \textrm{H}\textsc{i} clouds. The relatively large scatter in the EW$_{LIS}$ vs. EW$_{Ly\alpha}$ relation compared to the other two suggests the galaxy-to-galaxy variation when using LIS lines as a probe to trace \textrm{H}\textsc{i}, as we investigate in Section \ref{sec:covfrac}.
We discuss in detail the astrophysical picture suggested by our results in Section \ref{sec:models}.

\begin{table*}
\centering
\begin{threeparttable}
  \caption{Correlation Coefficients}
  \label{tab:corr}
{\renewcommand{\arraystretch}{1.5}
  \begin{tabular}{ccccc}
  \hline
  \hline
    Correlation & Kendall $\tau$ \tnote{1} & $P_{K}$\tnote{2} & Spearman $\rho$ \tnote{1} & $P_{SR}$\tnote{2} \\
  \hline
  \multicolumn{5}{c}{Entire Sample (157 objects)} \\
  \hline
  EW$_{LIS}$ vs. $E(B-V)$ & 5.301 $\pm$ 0.557 & 0.0000 $\pm$ 0.0000\tnote{3} & -0.396 $\pm$ 0.042 & 0.0000 $\pm$ 0.0007 \\ 
 \hline
 EW$_{LIS}$ vs. EW$_{Ly\alpha}$ & 3.520 $\pm$ 0.503 & 0.0004 $\pm$ 0.0066 & 0.273 $\pm$ 0.038 & 0.0006 $\pm$ 0.0090 \\ 
 \hline
 EW$_{Ly\alpha}$ vs. $E(B-V)$ & 3.859 $\pm$ 0.248 & 0.0001 $\pm$ 0.0001 & -0.299 $\pm$ 0.019 & 0.0002 $\pm$ 0.0002 \\ 
 \hline
 \multicolumn{5}{c}{Only LIS Detections (124 objects)\tnote{4}} \\
  \hline
  EW$_{LIS}$ vs. $E(B-V)$ & 6.340 $\pm$ 0.655 & 0.0000 $\pm$ 0.0000 & -0.531 $\pm$ 0.053 & 0.0000 $\pm$ 0.0000 \\ 
 \hline
 EW$_{LIS}$ vs. EW$_{Ly\alpha}$ & 3.862 $\pm$ 0.581 & 0.0001 $\pm$ 0.0074 & 0.345 $\pm$ 0.051 & 0.0001 $\pm$ 0.0082 \\ 
 \hline
 EW$_{Ly\alpha}$ vs. $E(B-V)$ & 4.505 $\pm$ 0.272 & 0.0000 $\pm$ 0.0000 & -0.391 $\pm$ 0.023 & 0.0000 $\pm$ 0.0000 \\  
\hline
 \end{tabular}}
\begin{tablenotes}
\item[1] Test statistic. The uncertainties on the correlation coefficients and P-values were derived by perturbing the measurements by their associated error bars. 
\item[2] Probability of a null hypothesis.
\item[3] A $P_{K}$ or $P_{SR}$ value or uncertainty listed as 0.0000 indicates an actual value less than $10^{-4}$ (but non-zero) and below the limit of precision offered by ASURV.
\item[4] The EW$_{LIS}$ detections include the `D', `P', and `C' objects, as defined in Section \ref{sec:lis}.
 \end{tablenotes}
\end{threeparttable}
\end{table*}

\section{Discussion}
\label{sec:discussion}

We have shown in Section \ref{sec:results} that EW$_{Ly\alpha}$, EW$_{LIS}$, and $E(B-V)$ are inter-correlated. While all three relations are statistically significant, the tightest correlation is found between EW$_{LIS}$ and $E(B-V)$. This particular result highlights the direct connection between dust and metal-enriched \textrm{H}\textsc{i} gas, suggesting that they are likely to be co-spatial. Additionally, the EW$_{Ly\alpha}$ vs. EW$_{LIS}$ correlation is found to be weaker than that between EW$_{LIS}$ and $E(B-V)$. This finding differs from the speculation made by multiple previous studies that EW$_{Ly\alpha}$ and EW$_{LIS}$ are the most directly connected \citep{Shapley2003,Du2018}. Although the strength of Ly$\alpha$ is directly modulated by the covering fraction of \textrm{H}\textsc{i} gas through resonant scattering, noticeable scatter can be introduced to the EW$_{Ly\alpha}$ vs. EW$_{LIS}$ relation by the object-to-object variation in the metal to \textrm{H}\textsc{i} covering fraction ratio (see a detailed discussion in Section \ref{sec:covfrac}). 
As detailed in Sections \ref{sec:clumpy} and \ref{sec:outflow}, we identify the dust content in between the \textrm{H}\textsc{i} gas clumps and outflow kinematics as two key contributors to the scatter in this relation. 

In this section, we review two ISM/CGM models involving the physical distributions of neutral hydrogen gas, dust, and metals. The empirical results presented in 
Section~\ref{sec:results} can be interpreted with reference to these ISM/CGM models, leading to insights into the distribution and kinematics of interstellar and circumgalactic gas, metals, and dust, and the escape of ionizing and Ly$\alpha$ photons. We then discuss the origins of the intrinsic scatter in the correlations observed among EW$_{Ly\alpha}$, EW$_{LIS}$, and $E(B-V)$.

\subsection{ISM/CGM Models}
\label{sec:models}

The physical picture underlying the observed trends among EW$_{Ly\alpha}$, EW$_{LIS}$, and $E(B-V)$ at $z\sim2-4$ has been considered in previous work  \citep[e.g.,][]{Shapley2003,Jones2012,Du2018}. In such descriptions, LIS absorption arises from metal-enriched clouds, which exist within a medium of patchy, neutral hydrogen gas. The ISM/CGM is considered porous due to the presence of outflows induced by active star formation commonly observed at high redshift \citep{Pettini2001,Shapley2003,Steidel2004}. While Ly$\alpha$ photons are resonantly scattered by the \textrm{H}\textsc{i} gas, they eventually escape through ``holes" in the \textrm{H}\textsc{i} gas where the \textrm{H}\textsc{i} column density or covering fraction is low, or by being back-scattered off of receding gas on the far side of the outflow. At the same time, we must consider the dust content of the absorbing \textrm{H}\textsc{i} gas in this picture, which is responsible for attenuating both the UV continuum and Ly$\alpha$ photons.

We do not yet have a clear picture, however, of the structure of the CGM and where dust resides with respect to the \textrm{H}\textsc{i} gas in the ISM/CGM of typical, star-forming galaxies at high redshift. As dust grains are formed by the condensation of metals, the distributions of metals and dust are expected to be highly correlated spatially.  Two basic CGM models that have been examined by previous studies \citep[e.g.,][]{Vasei2016,Gazagnes2018,Steidel2018} include: (1) a picket-fence \textrm{H}\textsc{i} gas model with a uniform foreground dust screen; and (2) a clumpy \textrm{H}\textsc{i} gas model where dust and metals are only located in the \textrm{H}\textsc{i} gas clumps. We note that in both models described above, the neutral hydrogen gas is always considered ``picket-fence-like" with a non-unity covering fraction, and the major difference lies in the distribution of dust and metals in the ISM/CGM. In this section, we compare our results with the model predictions and discuss which model our results are most likely to support. 

\subsubsection{Uniform Dust Screen Model}
\label{sec:screen}

As described in previous work \citep{Vasei2016,Gazagnes2018,Steidel2018}, the uniform dust screen model assumes that patchy \textrm{H}\textsc{i} gas in the ISM is dust- and metal-free, whereas dust, along with the metals that give rise to the low-ions, exists in a foreground, uniform screen.
If dust is uniformly distributed in and {\it only} in the foreground with a $100\%$ covering fraction, the attenuation affects both Ly$\alpha$ photons and the UV continuum to the same extent. The observed Ly$\alpha$ flux originates from the escaped Ly$\alpha$ photons, either directly from the Ly$\alpha$-emitting region or after multiple resonant scattering events, through channels of \textrm{H}\textsc{i} gas with low covering fraction or column density. In this work, we used EW$_{Ly\alpha}$ (instead of Ly$\alpha$ flux) as an observable, which is defined as the ratio of integrated Ly$\alpha$ flux and the continuum flux density redward of Ly$\alpha$. As the uniform dust screen attenuates the observed Ly$\alpha$ flux and the continuum to the same degree, the resulting EW$_{Ly\alpha}$ should be uncorrelated with $E(B-V)$ and only dependent on the \textrm{H}\textsc{i} covering fraction. Our results, however, contradict this prediction. We observe a significant anti-correlation between EW$_{Ly\alpha}$ and $E(B-V)$, where EW$_{Ly\alpha}$ decreases with increasing $E(B-V)$, 
which would not occur if dust only existed in a uniform foreground screen but not in the \textrm{H}\textsc{i} gas clumps.

In addition, the assumptions in the dust screen model suggest that the covering fractions of dust and \textrm{H}\textsc{i} gas are independent: the \textrm{H}\textsc{i} covering fraction can vary while the dust covering fraction is always $100\%$. Studies have shown that the metal covering fraction is positively correlated with the \textrm{H}\textsc{i} covering fraction \citep{Reddy2016,Gazagnes2018}, providing justification of using metal lines to probe \textrm{H}\textsc{i} gas. If the covering fractions of dust and \textrm{H}\textsc{i} gas were truly independent, EW$_{LIS}$ and $E(B-V)$ would display no apparent correlation. On the contrary, our findings show not only that EW$_{LIS}$ and $E(B-V)$ are correlated, but that their correlation is the strongest among the three key relations highlighted in this work (i.e., those connecting EW$_{Ly\alpha}$, EW$_{LIS}$, and $E(B-V)$). This strong connection between EW$_{LIS}$ and $E(B-V)$ points to the possibility that the metal-enriched \textrm{H}\textsc{i} gas and dust may in fact be co-spatial. Finally, we caution that the uniform dust screen model is not very probable in an astrophysical context. Provided that dust grains are formed from the condensation of metals, we expect the covering fractions and spatial distributions of neutral hydrogen gas, metals, and dust are related to some extent. It is extremely unlikely that dust exists in empty space or $only$ ionized gas but not in neutral hydrogen gas, such that the dust covering fraction is completely independent of the \textrm{H}\textsc{i} covering fraction.

\subsubsection{Dusty ISM/CGM Model}
\label{sec:clumpy}

We now consider the second model, where dust and metals are exclusively confined to \textrm{H}\textsc{i} gas clumps. In the simplest scenario, the intra-clump medium (ICM) is free of \textrm{H}\textsc{i} gas, dust, and metals. In this model, Ly$\alpha$ photons are scattered multiple times by the \textrm{H}\textsc{i} gas before escaping, leading to a longer path traveling through the ISM/CGM and higher probability of being attenuated by dust compared to the continuum photons. There are a few variations of the dusty ISM/CGM model, which have different predictions depending on specific assumptions. \citet{Neufeld1991} proposes a model where the dusty \textrm{H}\textsc{i} gas clumps are surrounded by an ICM that is optically thin to Ly$\alpha$ and has negligible dust content. In this model, the clumps are composed of optically-thick \textrm{H}\textsc{i} gas with high column density, and Ly$\alpha$ photons are scattered off of the clump surface without interacting with the shielded dust. Consequently, Ly$\alpha$ photons spend most of their time in the ICM bouncing between the clump surfaces, barely getting absorbed. In the meantime, continuum photons travel through the dusty clumps and are attenuated by the embedded dust. This model therefore predicts a positive correlation between EW$_{Ly\alpha}$ and $E(B-V)$ (as Ly$\alpha$ photons are {\it less} attenuated than the continuum photons by dust), and has been used to examine the ``boosting" of EW$_{Ly\alpha}$ in LAEs in high-redshift galaxies \citep{Hansen2006,Laursen2013,Duval2014}, where the observed EW$_{Ly\alpha}$ is larger than what is theoretically expected.

However, previous observations showing an anti-correlation between EW$_{Ly\alpha}$ and $E(B-V)$  \citep[e.g.,][]{Shapley2003,Pentericci2007,Marchi2019,Du2018, Steidel2011} are in conflict with the \citet{Neufeld1991} model. Furthermore, radiative transfer simulations have suggested that the \citet{Neufeld1991} model requires weak to no outflows and high density contrast between clumps and the ICM \citep{Laursen2013,Duval2014}, which do not appear to be realistic for the majority of galaxies at $z\gtrsim2$.

These discrepancies motivate us to seek alternative dusty ISM/CGM models. One possibility is that the clumpiness of the ISM/CGM has low contrast, such that the ICM is also optically thick to Ly$\alpha$ photons (but not as high density as the gas clumps) and has a non-negligible dust content \citep[the ``low-contrast" regime in][]{Duval2014}. In this scenario, Ly$\alpha$ photons cannot escape without scattering in the ISM/CGM because of the optically thick ICM. Hence, even if Ly$\alpha$ photons are still scattered at the clump surfaces and remain unaffected by the dust embedded within, Ly$\alpha$ flux can be more attenuated than the UV continuum flux because of the longer path Ly$\alpha$ photons take on average (due to resonant scattering) through the ICM compared to continuum photons before escaping. We can therefore observe a negative correlation between EW$_{Ly\alpha}$ and $E(B-V)$ as long as the cumulative $E(B-V)$ in the ICM at the point of escape for Ly$\alpha$ photons is larger than the $E(B-V)$ in the \textrm{H}\textsc{i} clumps. Another possibility is that the Ly$\alpha$ photons are in fact capable of penetrating the gas clumps and being absorbed by the embedded dust. This scenario requires the clumps to have a small velocity offset from the galaxy's systemic velocity, either in random motions or induced by stellar feedback, such that the Ly$\alpha$ photons are out of resonance with the \textrm{H}\textsc{i} gas clumps \citep{Hansen2006,Laursen2013}. The requirement of neutral clump motion is not difficult to fulfill at $z\sim2$, given the ubiquitously observed outflows in star-forming galaxies \citep{Shapley2003,Steidel2004,Du2018}. A scenario in which Ly$\alpha$ penetrates dusty clumps also leads to the differential attenuation between Ly$\alpha$ and the continuum photons, and predicts a negative correlation between EW$_{Ly\alpha}$ and $E(B-V)$ regardless of the dust content in the ICM.

Our results in Section \ref{sec:results} clearly agree more with a clumpy (and dusty) ISM/CGM model. Specifically, (1) EW$_{LIS}$ and $E(B-V)$ shows the strongest correlation among the three correlations we have examined, indicating that the covering fractions of \textrm{H}\textsc{i}, dust, and metals are tightly connected. This piece of observational evidence suggests the possibility of \textrm{H}\textsc{i}, dust, and metals being co-spatial, which is consistent with the dusty ISM/CGM model where dust and metals reside in the \textrm{H}\textsc{i} gas clumps; (2) EW$_{Ly\alpha}$ and $E(B-V)$ on average exhibit a negative correlation, consistent with Ly$\alpha$ photons within the region covered by the spectroscopic slit experiencing more attenuation than the continuum photons because of their longer paths through a dusty medium before escaping. An alternative explanation is that more Ly$\alpha$ photons are scattered out of the spectroscopic slit (while continuum photons remain unaffected) in galaxies with higher $E(B-V)$, which typically also have a higher \textrm{H}\textsc{i} covering fraction \citep{Reddy2016}. Either scenario described above requires the neutral hydrogen clumps to contain dust in order to explain the observed EW$_{Ly\alpha}$ vs. $E(B-V)$ trend; (3) Although the observed anti-correlation between EW$_{Ly\alpha}$ and $E(B-V)$ indicates that the \citet{Neufeld1991} scenario does not apply to the vast majority of the galaxies in our sample, we do observe a few outliers in the EW$_{Ly\alpha}$ vs. $E(B-V)$ relation (AEGIS-36451, COSMOS-2672, COSMOS-3974, COSMOS-6963, and COSMOS-26073). These galaxies are characterized by both relatively high global $E(B-V)$ and prominent Ly$\alpha$ emission, and objects with similar $E(B-V)$ and Ly$\alpha$ properties have been reported by previous studies \citep{Hagen2014,Trainor2016}. Interestingly, the presence of such outliers may indeed point to ``boosted" Ly$\alpha$ emission in these galaxies \citep{Neufeld1991}. For these outliers, Ly$\alpha$ photons 
are perhaps less attenuated than the continuum photons if they are simply scattered off of the surfaces of high column-density neutral clumps, and consequently become insensitive to the dust embedded within the clumps. In short, while the \citet{Neufeld1991} model cannot explain the overall EW$_{Ly\alpha}$ vs. $E(B-V)$ trend we observe, it may be applicable to the handful ``unusual" galaxies with both large EW$_{Ly\alpha}$ and high $E(B-V)$. All the results listed above are in support of the dusty ISM/CGM model, where dust and metals are located within neutral hydrogen gas clumps. While beyond the scope of this work, we note that our data allow variation in the dust-to-metals ratio, which partially contributes to the apparent intrinsic scatter in the EW$_{LIS}$ vs. $E(B-V)$ relation. Such variation has been shown from both observational and theoretical standpoints at low redshift \citep[e.g.,][]{Chiang2018,Li2019}, and may result from a large range of galactocentric radii where the LIS absorption takes place.

It is unfortunately impossible for us to determine the exact dust content in the ICM (e.g., dust-free or non-negligible dust) using the current LRIS-ISM data. Predicted observables (such as the emergent Ly$\alpha$ profile, and the expected strength or slope of the EW$_{Ly\alpha}$ vs. $E(B-V)$ relation) are needed to test the hypotheses and distinguish these cases. Future simulations studying the propagation of Ly$\alpha$ photons through the clumpy, dusty ISM/CGM with different density contrasts between clumps and the ICM will provide key insights into this question. We further attribute the ICM dust content partially to the observed scatter in the EW$_{Ly\alpha}$ vs. $E(B-V)$ relation. It is possible that the ICM dust varies on a galaxy-by-galaxy basis, depending on the detailed star-formation histories and stellar feedback of individual galaxies. We defer the discussion of the intrinsic scatter in the correlations to Section \ref{sec:scatter}.

\subsection{Intrinsic Scatter}
\label{sec:scatter}

Although statistically significant, the correlations among EW$_{Ly\alpha}$, EW$_{LIS}$, and $E(B-V)$ are all subject to non-negligible scatter introduced by different physical processes. To further investigate the origin of the intrinsic scatter in these relations, we test whether the observed galaxy-to-galaxy variation is correlated with any of the key galaxy properties under consideration.

\subsubsection{Scatter Introduced by Metallicity}
\label{sec:covfrac}

\begin{figure}
\includegraphics[width=1.0\linewidth]{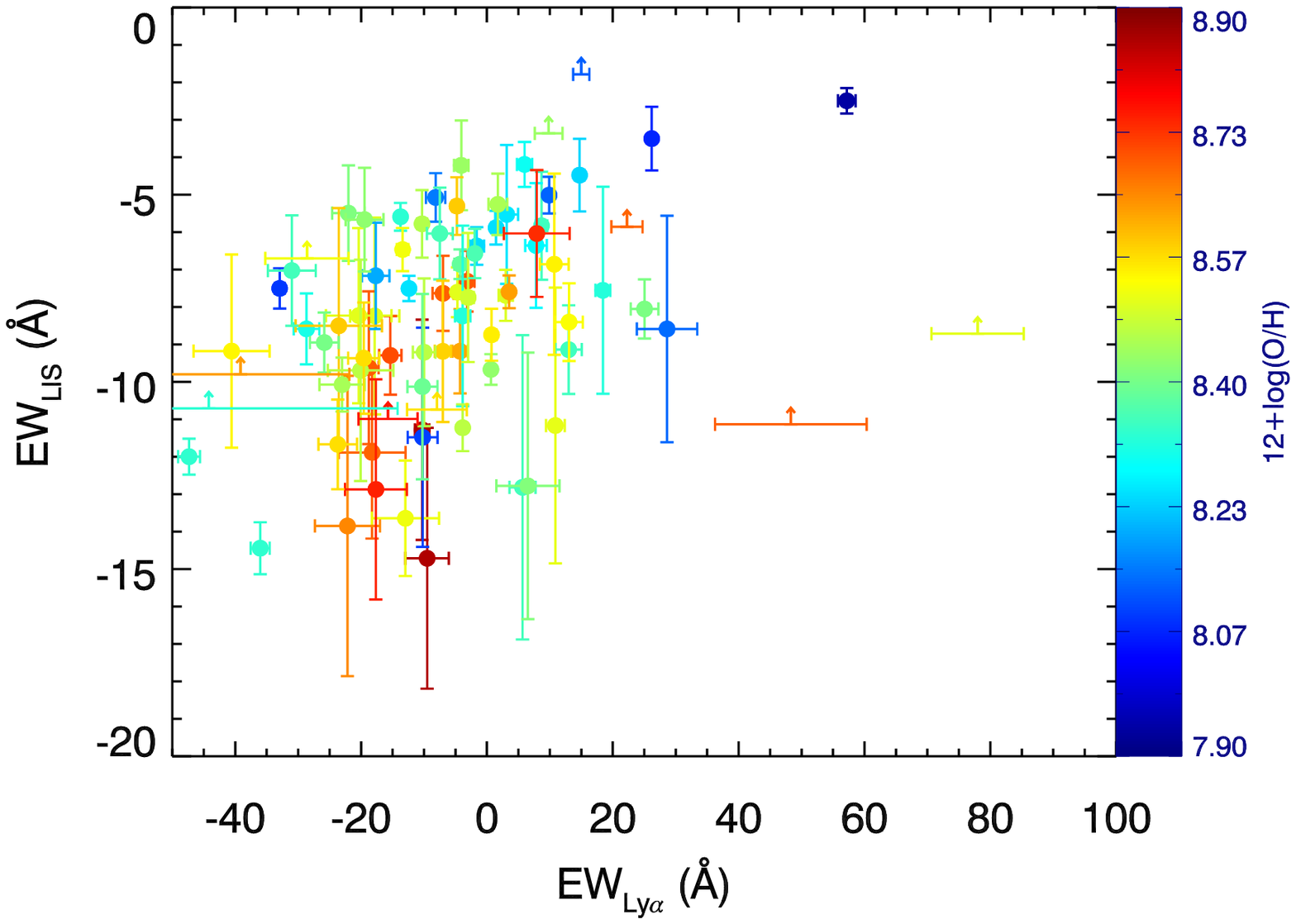}
\includegraphics[width=1.0\linewidth]{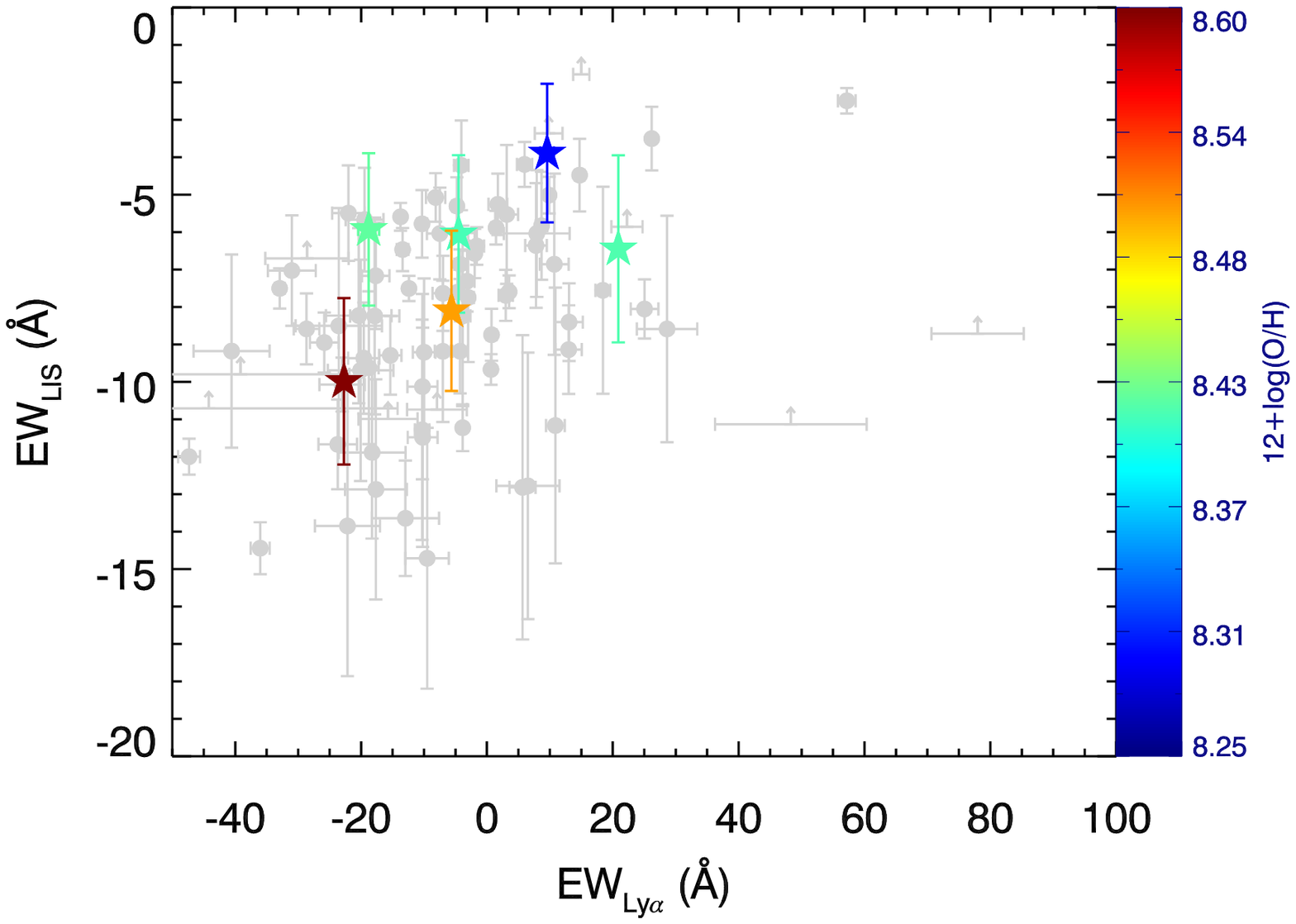}
\includegraphics[width=1.0\linewidth]{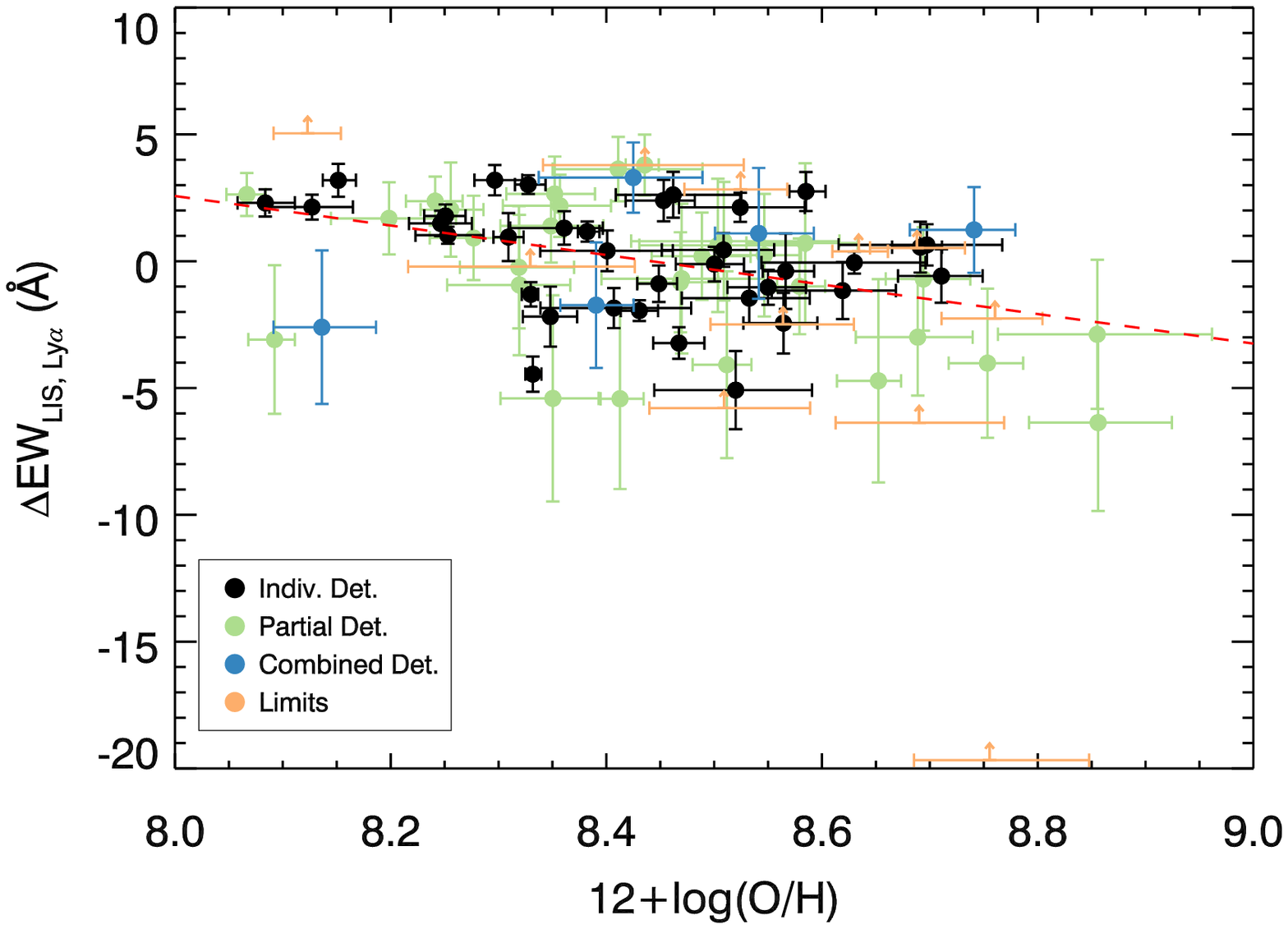}
\caption{The EW$_{LIS}$ vs. EW$_{Ly\alpha}$ relation as a function of metallicity. \textbf{Top:} EW$_{LIS}$ vs. EW$_{Ly\alpha}$ for 82 objects with individual metallicity measurements. The data points are color coded by metallicity. The 1$\sigma$ error bar is plotted for the LIS detections, while a 3$\sigma$ limit is shown as an upward-pointing arrow for the LIS non-detections. \textbf{Middle:} EW$_{LIS}$ vs. EW$_{Ly\alpha}$ for composite spectra created using 82 objects with individual metallicity measurements. The stacks were constructed by dividing 82 galaxies first into 3 bins in EW$_{Ly\alpha}$, and then into higher and lower halves in EW$_{LIS}$ in each EW$_{Ly\alpha}$ bin. The median metallicity in each stack is shown. The individual measurements of all 157 LRIS-ISM galaxies are plotted as gray points for comparison. \textbf{Bottom:} The ``residual" EW$_{LIS}$ at given EW$_{Ly\alpha}$ vs. metallicity. The ``residual" EW$_{LIS}$ was calculated by taking the difference of the measured EW$_{LIS}$ and the expected value at its EW$_{Ly\alpha}$, indicated by the mean EW$_{LIS}$ vs. EW$_{Ly\alpha}$ relation in Section \ref{sec:results}. The red dashed line denotes the best-fit linear regression.}
\label{fig:metal}
\end{figure}

Using EW$_{LIS}$ as a tracer of \textrm{H}\textsc{i} covering fraction is based on several assumptions, which individually can introduce uncertainties into the observed EW$_{LIS}$ vs. EW$_{Ly\alpha}$ relation. Metal absorption in the spectra of galaxies is often used as a proxy for the neutral hydrogen content of the ISM. As the LIS lines are saturated in the LRIS spectra analyzed in this work, EW$_{LIS}$ is not sensitive to the metal column density but instead the metal covering fraction. EW$_{Ly\alpha}$, on the other hand, is a probe of \textrm{H}\textsc{i} covering fraction. Although the metal covering fraction can be used as a proxy for the \textrm{H}\textsc{i} covering fraction, the former is found to be systematically smaller than the latter on average \citep{Reddy2016,Gazagnes2018} and the ratio between the two can vary on an individual basis, as described below.

To further explore whether the metal to \textrm{H}\textsc{i} covering fraction ratio is a potential source of scatter, we examine the role of metallicity in the EW$_{LIS}$ vs. EW$_{Ly\alpha}$ relation. As demonstrated in \citet{Gazagnes2018} using \textrm{Si}\textsc{ii} as a proxy for EW$_{LIS}$, one reason why the covering fraction of metals is correlated with, but not equal to, the \textrm{H}\textsc{i} covering fraction is that the metals are not fully mixed with the neutral hydrogen gas. These authors suggest that there are metal-enriched ``pockets" residing in the \textrm{H}\textsc{i} gas. Furthermore, in lower-metallicity galaxies, the metal-to-\textrm{H}\textsc{i} covering fraction ratio is lower because there is less metal-rich gas and therefore fewer high-density metal regions to absorb the background continuum.\footnote{Although \citet{Gazagnes2018} have shown that the covering fraction ratio between \textrm{H}\textsc{i} and \textrm{Si}\textsc{ii} has a weak dependence on metallicity, the deduced relationship is only marginally more significant than that assuming the covering fraction of \textrm{H}\textsc{i} solely depends on the \textrm{Si}\textsc{ii} covering fraction. Therefore, it is still reasonable to assume that the covering fractions of \textrm{H}\textsc{i} and metals are closely correlated, although additional scatter can be introduced by different chemical abundance patterns when using specific metal-line probes.} Accordingly, at fixed EW$_{Ly\alpha}$ (or approximately fixed \textrm{H}\textsc{i} covering fraction), we will observe lower-metallicity objects having smaller metal covering fractions, translating into weaker LIS absorption lines (lower EW$_{LIS}$) at fixed EW$_{Ly\alpha}$. Simulations such as that presented in \citet{Mauerhofer2021} will be helpful for testing this scenario.

In the top 2 panels of Figure \ref{fig:metal} we show measurements from individual galaxies (top) and composite spectra (middle), color coded by metallicity, in the EW$_{Ly\alpha}$ vs. EW$_{LIS}$ parameter space.  Metallicity for individual objects was determined based on O32$\equiv$[\textrm{O}\textsc{iii}]$\lambda5007$/[\textrm{O}\textsc{ii}]$\lambda\lambda3727,3729$ and other indicators when available, such as O3$\equiv$[\textrm{O}\textsc{iii}]$\lambda5007$/H$\beta$, and Ne3O2$\equiv$[\textrm{Ne}\textsc{iii}]$\lambda3869$/[\textrm{O}\textsc{ii}]$\lambda\lambda3727,3729$, as described in \citet{Sanders2020}. Such metallicity measurements make use of all available rest-optical emission lines from $\alpha$ elements (e.g., O, Ne). Using other common metallicity indicators, such as [\textrm{N}\textsc{ii}]$\lambda6584$/H$\alpha$ and O3N2$\equiv$([\textrm{O}\textsc{iii}]$\lambda5007$/H$\beta$)/([\textrm{N}\textsc{ii}]$\lambda6584$/H$\alpha$) \citep{Pettini2004}, yields the same qualitative trends presented below.

To obtain robust metallicity measurements, we required spectral coverage and a $S/N\geqslant3$ detection of at least [\textrm{O}\textsc{ii}] and [\textrm{O}\textsc{iii}], and of H$\beta$ and/or [\textrm{Ne}\textsc{iii}] when applicable. As a result, 82 out of 157 objects (52$\%$) have a valid metallicity estimate. We note that these 82 objects are representative of the full LRIS-ISM sample in galaxy properties, except that they all have redshift above $z=2.0857$ to allow coverage of [\textrm{O}\textsc{ii}] in the $J-$band. We used the same 82 objects to create composite spectra, first dividing the sub-sample into 3 bins in EW$_{Ly\alpha}$ and then higher- and lower-halves in EW$_{LIS}$ within each EW$_{Ly\alpha}$ bin (6 bins in total). Composite science and error spectra were created following the methodology of \citet{Du2018}. In short, we performed median stacking after interpolating individual galaxy spectra in each bin onto the same grid in wavelength. The corresponding composite error spectra were constructed by calculating the standard deviation of 100 fake median stacks at each wavelength, where the fake stacks were created by bootstrap resampling the objects in each bin, and perturbing individual galaxy spectra in the bootstrapped sample by their associated error spectra. The measurements of EW$_{Ly\alpha}$ and EW$_{LIS}$ in the composite spectra followed the descriptions in Section \ref{sec:measure}, and we plot the median metallicity in each composite. Both individual and composite spectra suggest that at fixed EW$_{Ly\alpha}$, LIS features are stronger in higher-metallicity objects or stacks. 

The bottom panel of Figure \ref{fig:metal} further supports the claim that metallicity plays a key role in the observed EW$_{Ly\alpha}$ vs. EW$_{LIS}$ relationship. The ``residual" EW$_{LIS}$, $\Delta$EW$_{LIS,Ly\alpha}$, was calculated by subtracting from each measured  EW$_{LIS}$ the ``expected" EW$_{LIS}$ based on the galaxy's EW$_{Ly\alpha}$, as defined by the mean EW$_{Ly\alpha}$ vs. EW$_{LIS}$ relation in Table \ref{tab:reg}. In the bottom panel, we observe that $\Delta$EW$_{LIS,Ly\alpha}$ decreases with increasing metallicity. This trend is statistically significant (Spearman's $\rho=-0.314$; probability of a null hypothesis is 0.0048), and again indicates that LIS lines become stronger, at fixed EW$_{Ly\alpha}$, with increasing metallicity. This finding not only supports the proposed physical scenario of inhomogeneous metal mixing, but also highlights one key source of the scatter in the EW$_{Ly\alpha}$ vs. EW$_{LIS}$ relation: metallicity. 

Our results here can be compared with those presented in \citet{Trainor2019}, who find that EW$_{Ly\alpha}$ corresponds to higher O3 at fixed EW$_{LIS}$. Given that O3 is anti-correlated with metallicity in the regimes probed in both studies, our findings are in qualitative agreement with those in \citet{Trainor2019}. In our work, we interpret the results as a metallicity variance across the EW$_{Ly\alpha}$ vs. EW$_{LIS}$ parameter space, which contributes scatter to the observed relation.

In summary, while interstellar metal absorption lines have proven to be a reasonable tracer of \textrm{H}\textsc{i} gas, intrinsic scatter in the EW$_{LIS}$ vs. EW$_{Ly\alpha}$ relation inevitably arises from the object-to-object variation in the metal to \textrm{H}\textsc{i} covering fraction ratio. Here we connect this ratio with gas-phase metallicity, which is determined by the star-formation histories, stellar populations, galaxy age, and galactic-scale gas flows \citep[e.g.,][]{Tremonti2004,Sanders2020}. Other factors related to metallicity, such as nebular ionization, may further contribute to the intrinsic scatter in the in the EW$_{Ly\alpha}$ vs. EW$_{LIS}$ relation, as explored in previous work \citep{Trainor2019}.

\subsubsection{Scatter Introduced by Outflow Kinematics}
\label{sec:outflow}

\begin{figure*}
\includegraphics[width=0.5\linewidth]{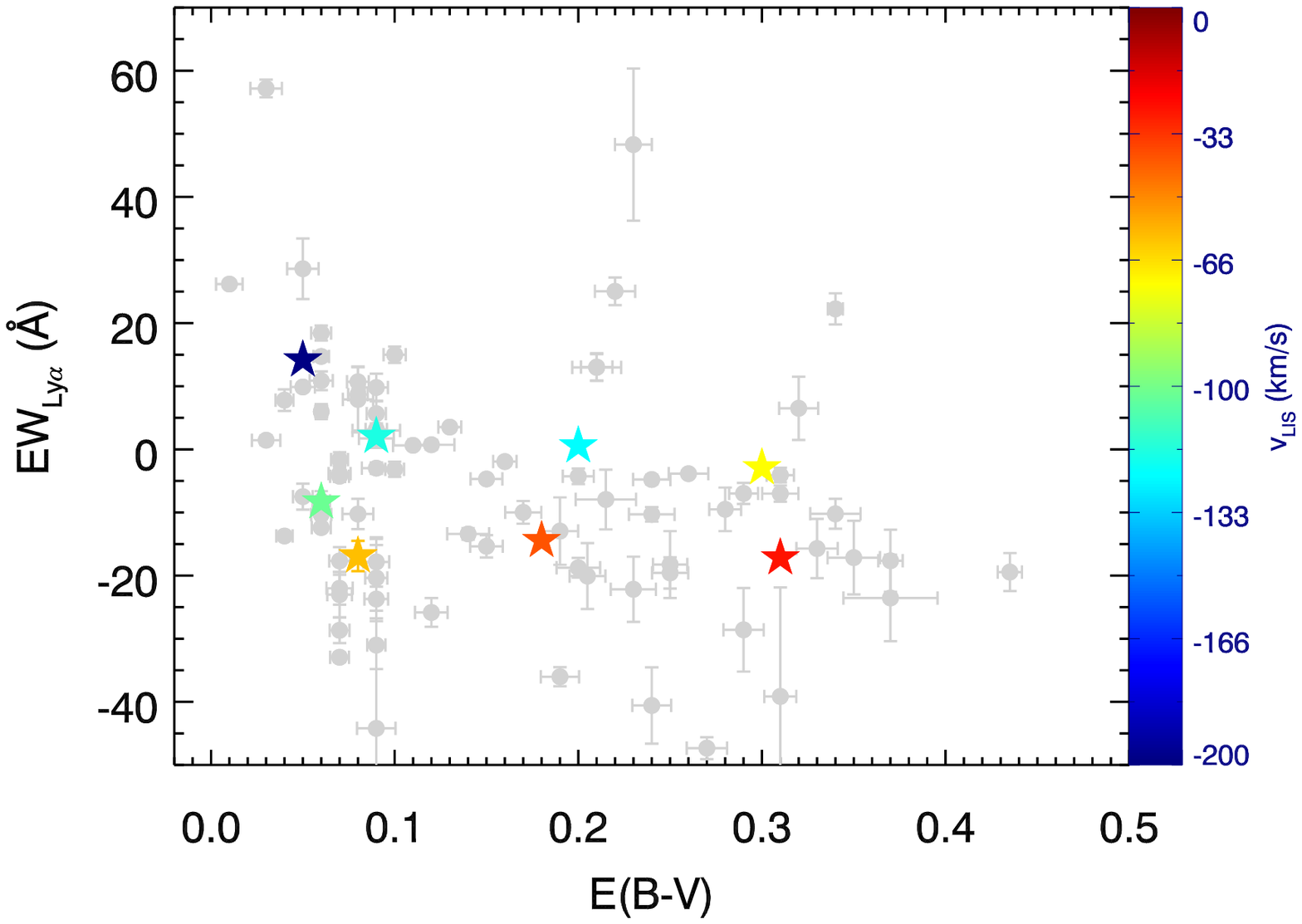}
\includegraphics[width=0.5\linewidth]{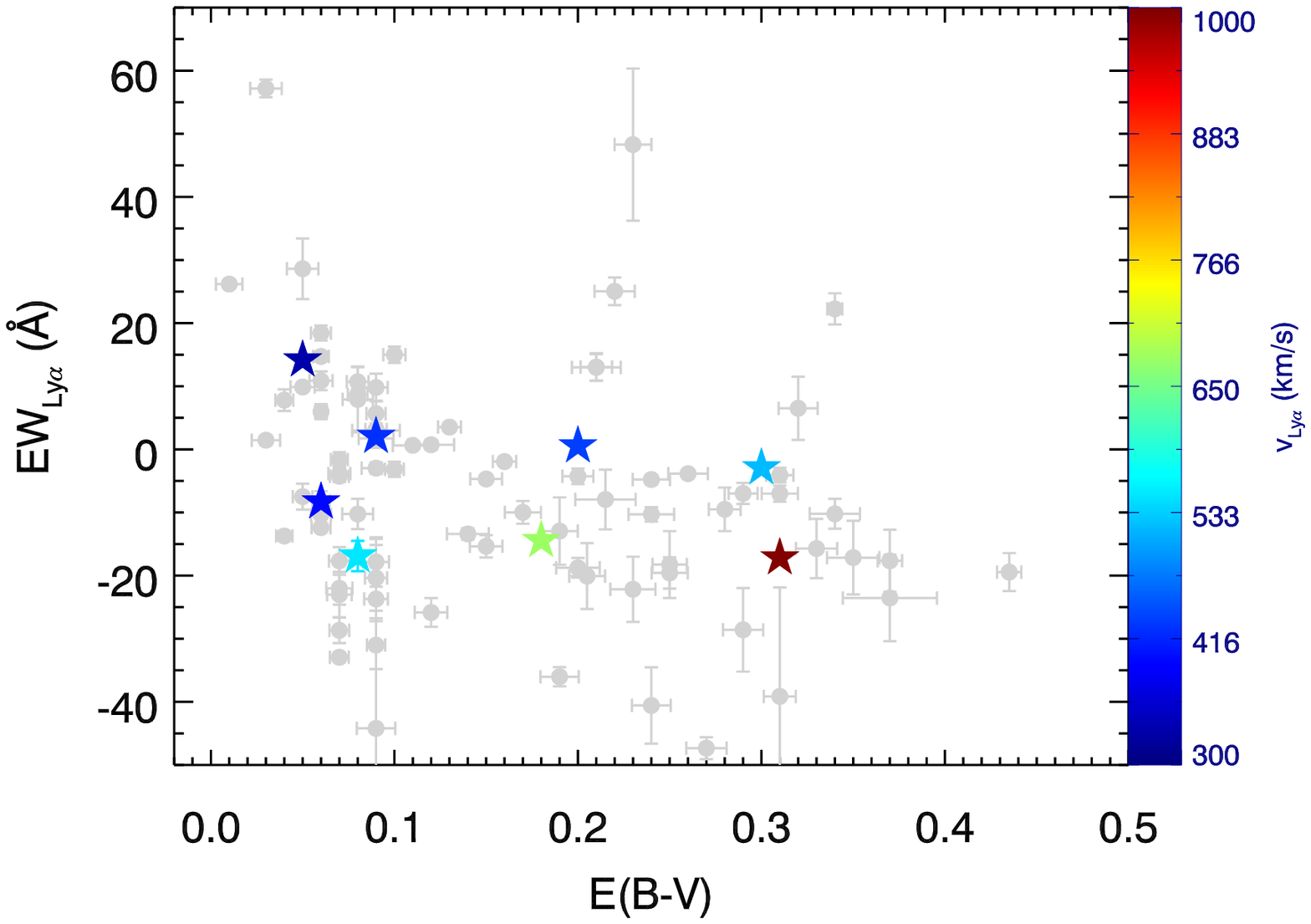}
\caption{EW$_{Ly\alpha}$ vs. $E(B-V)$ for composite spectra color coded by the LIS (left; $v_{LIS}$) and Ly$\alpha$ velocity shifts (right; $v_{Ly\alpha}$). The composite were constructed by dividing all 157 galaxies first into 4 bins of $E(B-V)$, and then into higher and lower halves in EW$_{Ly\alpha}$ in each $E(B-V)$ quartile. The individual measurements are plotted as gray points for comparison. The velocity shifts were calculated based on the average centroid velocity of \textrm{Si}\textsc{ii}$\lambda1260$,  \textrm{O}\textsc{i}$\lambda1302$+\textrm{Si}\textsc{ii}$\lambda1304$, and \textrm{C}\textsc{ii}$\lambda1334$ for $v_{LIS}$ and the centroid of the Ly$\alpha$ emission peak for $v_{Ly\alpha}$.}
\label{fig:vshift}
\end{figure*}

Galactic-scale outflows are ubiquitously observed at $z\gtrsim2$ \citep{Pettini2002,Shapley2003,Steidel2010} and have significant impact on the radiative transfer of Ly$\alpha$ photons through the neutral ISM/CGM. As a result, outflow kinematics can modulate the emergent Ly$\alpha$ profile and, accordingly, EW$_{Ly\alpha}$.

As detailed in Section \ref{sec:lya}, the Ly$\alpha$ feature in the LRIS-ISM spectra is observed to have various profile morphologies. Previous studies have examined how outflow kinematics can alter the observed Ly$\alpha$ profile. For example, \citet{Verhamme2006} used a 3D Ly$\alpha$ radiation transfer code to study the emergent Ly$\alpha$ line profiles in galaxy environments with different \textrm{H}\textsc{i} densities, dust distributions, and velocity fields. Additionally, \citet{Steidel2010} proposed an analytical model of the outflowing ISM/CGM to explain the variation in the observed Ly$\alpha$ emission profile. Their results suggest that the \textrm{H}\textsc{i} covering fraction near the galaxy's systemic velocity ($v\simeq0$) is responsible for the observed redshifted peak of Ly$\alpha$: the higher the \textrm{H}\textsc{i} covering fraction, the more absorption at $v\simeq0$, pushing the observed Ly$\alpha$ peak towards redder wavelength and reducing the overall EW$_{Ly\alpha}$. In terms of a physical picture, redder (and typically weaker) Ly$\alpha$ emission signals the fact that only photons emitted or scattered by the materials on the far side of the galaxy that have a redshifted velocity large enough to make Ly$\alpha$ photons out of resonance with the foreground \textrm{H}\textsc{i} can escape in the observer’s direction.

Given that outflow kinematics may be one factor modulating EW$_{Ly\alpha}$, we examine how Ly$\alpha$ and LIS velocity shifts ($v_{Ly\alpha}$ and $v_{LIS}$, respectively) affect the EW$_{Ly\alpha}$ vs. $E(B-V)$ relation. The velocity shift of a line is defined as $v=(\lambda_{obs}-\lambda_{rest})/\lambda_{rest}\times$$c$, where $\lambda_{obs}$, $\lambda_{rest}$ are, respectively, the observed and rest-frame wavelengths of the spectral feature. Figure \ref{fig:vshift} shows composite spectra, color coded by $v_{LIS}$ (left) and $v_{Ly\alpha}$ (right), in the EW$_{Ly\alpha}$ vs. $E(B-V)$ plane. The composites make use of all 157 objects in the LRIS-ISM sample, and were binned first according $E(B-V)$ and then divided into higher and lower halves in EW$_{Ly\alpha}$ in each $E(B-V)$ quartile. $v_{LIS}$ was estimated as the average centroid velocity of \textrm{Si}\textsc{ii}$\lambda1260$,  \textrm{O}\textsc{i}$\lambda1302$+\textrm{Si}\textsc{ii}$\lambda1304$, and \textrm{C}\textsc{ii}$\lambda1334$, where the centroid of respective lines was returned by MPFIT (see Section \ref{sec:lis}). We measured $v_{Ly\alpha}$ based on the Ly$\alpha$ emission peak, which was fit with a Gaussian profile over the wavelength range bracketed by the blue- to red-bases of the Ly$\alpha$ emission profile. We note that while weak, the Ly$\alpha$ emission peak is still discernible in stacks categorized as ``absorption" in Ly$\alpha$ morphology. 

We find that at fixed $E(B-V)$, Ly$\alpha$ is stronger in galaxies with larger LIS blueshifts and smaller Ly$\alpha$ redshifts. This result not only identifies outflow kinematics as a contributor to the observed scatter in the EW$_{Ly\alpha}$ vs. $E(B-V)$ relation, but also highlights conditions favorable for Ly$\alpha$ photon escape. The centroid velocity of LIS lines describes the bulk movement of the neutral, metal-enriched gas. Therefore, larger LIS blueshifts in general correspond to higher outflow velocities, with which outflows may clear out channels through the ISM/CGM efficiently, reducing the covering fraction of \textrm{H}\textsc{i} gas and allowing for the escape of Ly$\alpha$ photons. On the other hand, the higher the bulk outflow velocities, the smaller fraction of \textrm{H}\textsc{i} gas is expected to be moving at $v\simeq0$, resulting in less redshifted and stronger Ly$\alpha$ emission. Our findings agree with the predictions of the \citet{Steidel2010} model, and demonstrate that outflow kinematics, as determined by star-formation activities, can introduce considerable scatter in the observed EW$_{Ly\alpha}$ vs. $E(B-V)$ relation.

\subsubsection{Other Physical Origins of Scatter}
\label{sec:origin}

Aside from the key observables investigated in the previous sections, such as metallicity,  $v_{Ly\alpha}$, and $v_{LIS}$, other factors or physical processes may have contributed to the observed scatter in the correlations involving EW$_{Ly\alpha}$, EW$_{LIS}$, and $E(B-V)$. In this section, we briefly discuss the other possible origins of scatter in an astrophysical context.

$Ly\alpha$ $Production$: In Section \ref{sec:results}, the overall trend between EW$_{Ly\alpha}$ and EW$_{LIS}$ highlights the importance of resonant scattering of Ly$\alpha$ photons by \textrm{H}\textsc{i} gas. It is worth mentioning that the observed Ly$\alpha$ EW depends on not only the escape but also the $production$ of Ly$\alpha$ photons. Although this effect impacts all objects regardless of their EW$_{Ly\alpha}$, it is especially prominent in Ly$\alpha$ emitters (LAEs; rest-frame $EW_{Ly\alpha}\geqslant20\mbox{\AA}$), where the hard ionizing spectrum associated with metal-poor star formation not only boosts the intrinsic Ly$\alpha$ production, but also lowers the \textrm{H}\textsc{i} covering fraction by ionizing the \textrm{H}\textsc{i} gas in the ISM \citep{Trainor2016,Erb2016, Reddy2016}. While we expect the variation in Ly$\alpha$ production to play a relatively small role in the majority of the galaxies in the LRIS-ISM sample (the median EW$_{Ly\alpha}$ is $-6.0\mbox{\AA}$), there are in fact 15 galaxies with $EW_{Ly\alpha}\geqslant20\mbox{\AA}$. Hence, the $\sim10\%$ LAEs in the sample may have introduced some scatter in the observed relations involving EW$_{Ly\alpha}$, due to their higher than average Ly$\alpha$ production efficiency.

$Slit$ $Loss$: Slit spectroscopy is typically used for observing the compact continuum-emitting regions of high-redshift star-forming galaxies. At $z\sim2.3$, the median redshift of the LRIS-ISM sample, the slit width of $1.''2$ corresponds to a physical size of $\sim10$ kpc. As shown by previous work \citep{Steidel2010,Matsuda2012}, the cool-phase CGM, where low-ions are used to trace \textrm{H}\textsc{i} gas, has a physical scale $\gtrsim100$ kpc at $z\sim2-3$. Accordingly, slit spectra can only capture the CGM at relatively small galactocentric radii, and the inferred galaxy properties and measured line strengths are largely ``local" rather than ``global." In particular, multiple studies have reported extended Ly$\alpha$ halos ($\sim80-100$ kpc) surrounding star-forming galaxies at $z\sim2-3$ \citep{Steidel2011,Matsuda2012}. If the slit spectra can only collect Ly$\alpha$ photons emitted/escaping from the innermost 10 kpc, the observed EW$_{Ly\alpha}$ may not be an accurate proxy for the actual EW$_{Ly\alpha}$ we would observe with slitless spectroscopy.

To test the effect slit loss has on the variation of EW$_{Ly\alpha}$, we obtained the continuum size of the MOSDEF-LRIS galaxies in the HST/F160W filter, drawn from the \citet{vanderWel2014} catalog. Preliminary results suggest that $\Delta$EW$_{Ly\alpha,E(B-V)}$ (i.e., the ``residual" EW$_{Ly\alpha}$ at fixed $E(B-V)$ as predicted by the mean EW$_{Ly\alpha}$ vs. $E(B-V)$ relation) is neither correlated with the galaxy physical size (in kpc) nor the ratio of the slit width and the galaxy angular size in the $H-$band. We caution that our result is inconclusive in determining the contribution of slit loss to the scatter in the EW$_{Ly\alpha}$ vs. $(B-V)$ relation, as the galaxy size indicated by stellar continuum emission is different from that indicated by Ly$\alpha$ emission. Hence, narrow-band imaging and integral field unit (IFU) observations (e.g., the Keck Cosmic Web Imager) are needed to (1) characterize the physical scale of the Ly$\alpha$ halo in these galaxies and quantify the effect of slit loss on the measurement of EW$_{Ly\alpha}$; and (2) determine whether the observed anti-correlation between EW$_{Ly\alpha}$ vs. $E(B-V)$ originates from the preferential dust attenuation of Ly$\alpha$ photons relative to continuum photons or is simply an effect of the scattering of Ly$\alpha$ photons by \textrm{H}\textsc{i} out of the spectroscopic slit.

\section{Summary}
\label{sec:summary}

Rest-UV spectra provide unique information on the structure of the ISM/CGM in star-forming galaxies. The LIS absorption features are commonly used to probe the neutral CGM. Along with Ly$\alpha$ and other key galaxy properties such as dust extinction, the geometry and physical distributions of \textrm{H}\textsc{i} gas, metals, and dust can be inferred. Using deep rest-UV LRIS spectra, we obtained individual measurements of EW$_{Ly\alpha}$, EW$_{LIS}$, and $E(B-V)$ for a statistical sample of 157 objects. Rest-optical MOSFIRE spectra provide additional data for each object, including the systemic redshift and gas-phase metallicity. We characterize the tightness of the mutual correlations among EW$_{Ly\alpha}$, EW$_{LIS}$, and $E(B-V)$ using statistical analysis, and identify the factors that contribute to the intrinsic scatter in the relations. Below we summarize our main findings.

1. EW$_{Ly\alpha}$, EW$_{LIS}$, and $E(B-V)$ are found to be inter-correlated. EW$_{LIS}$ and $E(B-V)$ displays a positive correlation, and galaxies with stronger EW$_{LIS}$ or larger $E(B-V)$ on average have smaller EW$_{Ly\alpha}$. This result agrees with those from previous studies using measurements from composite galaxy spectra. This finding supports a picture where Ly$\alpha$ photons are resonantly scattered by the clumpy \textrm{H}\textsc{i} gas, being absorbed by dust as they travel through the ISM/CGM, and escape the ISM/CGM through channels with low \textrm{H}\textsc{i} covering fractions and/or column densities.

2. Using statistical analysis, we find that all three mutual correlations among EW$_{Ly\alpha}$, EW$_{LIS}$, and $E(B-V)$ are statistically significant. The strongest relation is between EW$_{LIS}$ and $E(B-V)$, while the EW$_{Ly\alpha}$ vs. EW$_{LIS}$ relation is the weakest. The ordering of correlation strengths does not depend the inclusion of EW$_{LIS}$ non-detections.

3. The fact that the EW$_{LIS}$ vs. $E(B-V)$ relation appears to be the most fundamental one among the three correlations highlights the physical connections between dust and metal-enriched \textrm{H}\textsc{i} gas, suggesting that they are likely to be co-spatial. The observed correlation between EW$_{Ly\alpha}$ and $E(B-V)$ suggests either the preferential dust attenuation of Ly$\alpha$ photons compared to the continuum photons, or a larger fraction of Ly$\alpha$ photons being scattered out of the slit in galaxies with higher $E(B-V)$ and \textrm{H}\textsc{i} covering fraction. Finally, part of the apparent scatter in the EW$_{Ly\alpha}$ vs. EW$_{LIS}$ relation can be explained by the difference in the metal-to-\textrm{H}\textsc{i} covering fraction ratio that is driven by variations in metallicity (as traced by rest-optical nebular-emission line ratios in our frame work) on an individual basis. The metal covering fraction, which determines the strength of the saturated LIS lines, does not directly probe the covering fraction of \textrm{H}\textsc{i} gas that modulates EW$_{Ly\alpha}$. We have further identified outflow kinematics and the amount of dust in the ICM as two factors that can contribute to the scatter in the EW$_{Ly\alpha}$ vs. $E(B-V)$ relation. 

4. In addition to metallicity and outflow kinematics, we have qualitatively determined other possible origins of the scatter in the mutual correlations involving EW$_{Ly\alpha}$, EW$_{LIS}$, and $E(B-V)$. These include (i) Ly$\alpha$ production efficiency, which is particularly important for LAEs \citep{Reddy2016}, where the emergent Ly$\alpha$ flux is not only determined by the escape but also the production of Ly$\alpha$ photons; and (ii) slit loss, which can potentially impact the global measurements of EW$_{Ly\alpha}$, EW$_{LIS}$, and $E(B-V)$ due to finite slit width. 

5. We have reviewed two previously proposed CGM models that consider different spatial distributions of \textrm{H}\textsc{i} gas, dust and metals. Based on the strongest correlation observed between EW$_{LIS}$ and $E(B-V)$, and the existence of objects with prominent Ly$\alpha$ emission and large $E(B-V)$, our data prefer the dusty ISM/CGM model where dust resides in the \textrm{H}\textsc{i} gas clumps instead of being distributed in a uniform foreground screen. The uniform dust screen model is further disfavored because it is physically unlikely and predicts no dependence between EW$_{Ly\alpha}$ and $E(B-V)$, which contradicts our findings.

Confirming the detailed ISM/CGM structure in typical star-forming galaxies requires not only high-quality multidimensional data, but also simulations that incorporate radiative transfer of Ly$\alpha$ and stellar feedback to compare with observational constraints. For example, narrow-band images focused near Ly$\alpha$ yield valuable information on the size of the Ly$\alpha$ halo and the extended CGM, providing calibrations and correction factors for the observed line EWs in slit spectroscopy. IFU spectroscopic maps will provide additional insights into the spatial variation of galaxy properties and line strengths, revealing the small-scale physical processes that lead to the intrinsic scatter in the observed correlations and informing CGM models based on the spatially-resolved gas and dust properties \citep[e.g.,][]{Bridge2018}. Finally, simulations with prescriptions of stellar feedback and radiative transfer can test different CGM models and predict the expected EW$_{Ly\alpha}$ and $E(B-V)$ relation based on different relative spatial distributions of \textrm{H}\textsc{i} gas and dust. A robust understanding of the structure of the CGM will enlighten us on topics such as (1) the escape fractions of Ly$\alpha$ and LyC photons; and (2) the nature of the relations among neutral hydrogen gas, dust, and metals. Further observational data and analytical models that directly address these questions are essential for making progress towards that end.

\acknowledgements 
We thank the referee, Ryan Trainor, for a thorough and constructive report, which improved the paper. We acknowledge support from NSF AAG grants AST1312780, 1312547, 1312764, and 1313171, grant AR13907 from the Space Telescope Science Institute, and grant NNX16AF54G from the NASA ADAP program. We also acknowledge a NASA contract supporting the ``{\it WFIRST} Extragalactic Potential Observations (EXPO) Science Investigation Team" (15-WFIRST15-0004), administered by GSFC. XD and BM further acknowledge support from the NASA MUREP Institutional Research Opportunity (MIRO) through the grant NNX15AP99A as well as from the MUREP Aerospace Academy (MAA) through the grant 80NSSC19M0099. We are grateful to the 3D-HST team for providing ancillary data on galaxy properties. We wish to extend special thanks to those of Hawaiian ancestry on whose sacred mountain we are privileged to be guests. Without their generous hospitality, most of the observations presented herein would not have been possible.

\bibliographystyle{apj}
\bibliography{ms}
\end{CJK}
\end{document}